\documentclass[useAMS,usenatbib,usegraphicx]{mn2e}

\usepackage{aas_macros}

\title[Mass segregation and elongation of Westerlund 1]{Mass segregation and elongation of the starburst cluster Westerlund 1  \thanks{Based on observations collected at the European Southern Observatory, La Silla, Chile, and retrieved from the ESO archive (Program ID 67.C-0514) }}
\author[M. Gennaro et al.]{M. Gennaro$^{1}$\thanks{E-mail:
gennaro@mpia.de (MG)}\thanks{Member of the International Max Planck Research School for Astronomy and Cosmic Physics at the University of Heidelberg, IMPRS-HD, Germany}, W. Brandner$^{1}$, A. Stolte$^{2}$, Th. Henning$^{1}$\\
$^{1}$Max-Planck-Institut f{\"u}r Astronomie, K\"{o}nigstuhl 17, 69117, Heidelberg, Germany\\
$^{2}$Argelander Institut f{\"u}r Astronomie, Auf dem H{\"u}gel 71, 53121, Bonn, Germany
}

\begin{document}

\date{Submitted to MNRAS 2010 October 14. Accepted 2010 November 23.}

\pagerange{\pageref{firstpage}--\pageref{lastpage}} \pubyear{2010}

\maketitle

\label{firstpage}

\begin{abstract}
Massive stellar clusters are  the best available laboratories to study the mass function of stars. Based on NTT/SofI near-infrared photometry, we have investigated the properties of the massive young cluster Westerlund 1.
From comparison with stellar models, we derived an extinction $A_{K_{\rmn{S}}} = 0.91 \pm 0.05$ mag, an age $\tau = 4 \pm 0.5$ Myr and a distance $d = 4.0 \pm 0.2$ kpc for Westerlund 1, as well as a total mass of $M_{\rmn{Wd} \,1} = 4.91_{-0.49}^{+1.79}  \times 10^4 M_{\sun}$.
Using spatially dependent completeness corrections we performed a 2D study of the cluster's IMF and, in addition, of the stellar density profiles of the cluster as a function of mass. From both IMF slope variations and stellar density, we find strong evidence of mass segregation.
For a cluster with some $10^5$ stars, this is not expected at such a young age as the result of two-body relaxation alone. We also confirm previous findings on the elongation of Westerlund 1; assuming an elliptical density profile, we found an axis ratio of a:b = 3:2.
Rapid mass segregation and elongation could be well explained as the results of subclusters merging during the formation of Westerlund 1.

 \end{abstract}


\section{Introduction}
Westerlund 1 (Wd~1) is among the most massive young clusters in the Local Group. Recent studies have revived interest in this cluster, discovered already 50 years ago \citep{1961PASP...73...51W}. Several of these studies focus on the rich population of massive stars that are spectroscopically identified as Wd~1 members \citep[see e.g.][]{Clark:2005lr,Negueruela:2005lr,Crowther:2006fk,Negueruela:2010fj}. Among this population it has been possible to find Wolf-Rayet stars, evolved OB stars, and short-lived transitional objects like Luminous Blue Variables and Yellow Hypergiants. Wd~1 is the only case in which such a rich population of these very rare objects is observable. This makes Wd~1 one of the most important templates for understanding the evolution of very massive stars after they leave the main sequence. One of  the great advantages is that the progenitor's mass of the evolved massive stars can be inferred from the observable main-sequence turn-off.
While the massive stars are bright enough to be observed at optical wavelengths, observations of the intermediate- and low-mass stellar population  is best performed in the near-infrared, given the high extinction towards Wd~1 of $A_V \sim 10 - 12 \,\rmn{mag}$ \citep*[][]{Piatti:1998qy}.
A study of the Wd~1 population below $\sim 30 M_{\sun}$ has been recently carried out by \cite{Brandner:2008lr} --hereafter Paper I--. In this paper we present comprehensive analysis of the data described in Paper I.

%Different approaches and different data sets give mass and age estimates for the Wd1 cluster that are in slight disagreement among each other; we will comment on this differences and their possible explanations in more details in the following. For the moment it is enough to say that, given
With the present estimates of its mass --$ \rmn{from}\, 5 \times 10^4 \, \rmn{to} \, 1.5\times 10^5 M_{\sun}$-- and age --from 3 to 6 Myr-- \citep[see][]{Clark:2005lr,Crowther:2006fk,Brandner:2008lr,Mengel:2009uq,Negueruela:2010fj}, Wd~1 represents probably the best template in the Milky Way to understand the cluster mode of star formation that can be observed in other galaxies, like the Antennae Galaxies, where Super Star Clusters with masses larger $ 10^5 M_{\sun}$ have been  detected  \citep[see e.g.][]{2010AJ....140...75W}. 

In addition to the study of the intriguing formation scenario of such massive extragalactic clusters, Wd~1 may also serve as a template to understand the interplay between evolution of massive stars and dynamical processes that may lead to the formation of stable, bound and relaxed globular clusters.
Given its mass, Wd~1 may indeed be able to retain a substantial fraction of its initial stellar population, even though, according to \cite{2006ApJ...650..203M}, it has probably undergone $\sim 65$ supernova events. These, in addition to stellar winds and ionizing radiation from the most massive stars, have dispersed the residual gas reservoir of the cluster, decreasing the gravitational binding energy of the system. If massive enough to resist disruption, Wd~1 will eventually turn into a closed, virialized system.
A study of the dynamical status of Wd~1 has been made by \cite{Mengel:2009uq}. The authors use the measured radial velocity dispersion for a group of 10 massive stars to infer a dynamical mass of $1.5 \times 10^5 M_{\sun}$, on the upper end of the Wd~1 mass estimates available in the literature. To derive this number, the authors assume virial equilibrium and isotropy of the stellar motions hence their estimate is an upper limit.
The analysis of star clusters' dynamical and structural parameters often assumes spherical symmetry. Hence, the cluster properties, like the IMF slope,  the stellar density profiles, the stars' velocity distributions, are described as 1D functions depending on the distance from the centre of the cluster. However the spherical symmetry assumption may not be valid, and this is the case of Wd~1. 
Several studies have already shown that Wd~1 has indeed an elongated shape, based on X-Ray diffuse emission \citep[][]{2006ApJ...650..203M} and stellar counts \citep[][]{Brandner:2008lr}.
Therefore, an unbiased study, which does not assume \emph{a priori} any symmetry for the geometry of Wd~1, is needed to properly investigate the spatial properties of the cluster.

We focus our attention on the study of mass segregation, global and spatially varying IMF  and overall cluster shape as can be derived by the study of density profiles. These macroscopic properties are, in turn, related to the formation history of the cluster, its internal dynamical evolution and its global interactions with the rest of the Galaxy.
We developed new analysis techniques to take into account the observational biases related to the presence of many very bright objects that can hamper a quantitative determination of both the IMF slope and the stellar density profiles.
The most important improvement compared to Paper I is that we drop any spherical symmetry assumption, regarding the cluster structure. Hence the completeness maps, the photometric errors and the density profiles are all obtained in a 2D approach. In addition, new stellar evolutionary models are used for comparison with observations. A probabilistic approach is developed to determine cluster memberships, using a nearby off-cluster image as a control frame for the field population.
Stellar masses are  derived using a maximum likelihood technique, taking into account realistic photometric errors and their correlations. IMF slopes are inferred using an approach which does not require any binning but makes use of all the information contained in each star's mass-probability-density-distribution.
We use 2D-elliptical-generalization of the radial density profiles by \cite*{Elson:1987lr} to obtain shape properties of Wd~1 (e.g. its semimajor axis, elongation and orientation).

The paper structure is as follows: we describe the data set used in Sect. \ref{sec:data}.
The technique to build completeness maps is introduced in Sect. \ref{sec:2dmaps}. In Sect. \ref{sec:everr} we use simulated stars to obtain photometric errors and their correlation.
A statistical field subtraction method is introduced in Sect. \ref{sec:subfield}. After the description of the adopted stellar models (Sect. \ref{sec:stellarmodels}), we use them and the clean Colour Magnitude Diagram of Wd~1 to infer its properties like extinction, age and distance (Sect. \ref{sec:fundpar}).
An approach to obtain, for each star, its mass-probability-distribution (given the adopted models) is shown in Sect. \ref{sec:IMF} where we also derive the global IMF slope and the variation of the IMF slope across the cluster.
In Sect. \ref{sec:morph} we build cluster density profiles and analyze them using elliptical models. We also quantify the extent of mass segregation.
The last section deals with our conclusions.
\begin{figure}
 \centering
\resizebox{\hsize}{!}{\includegraphics{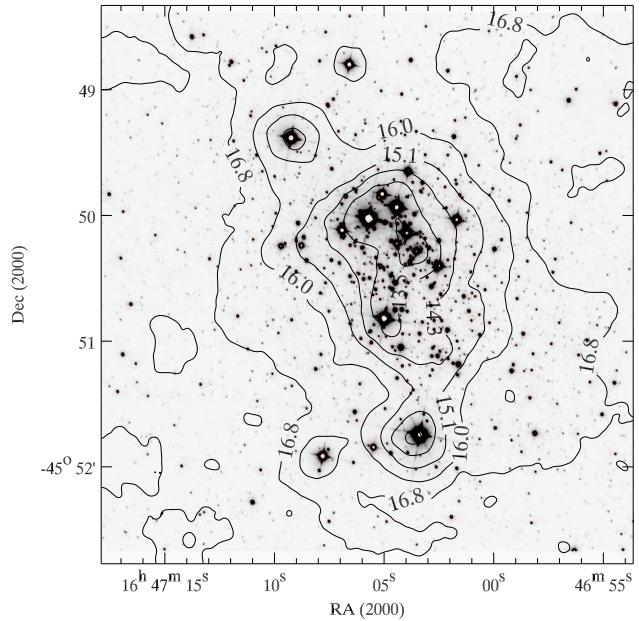}}
 \caption{SOFI $K_{\rmn{S}}$ band image of Wd~1. Superimposed are $K_{\rmn{S}}$ 50\% completeness contours. The labels correspond to the $K_{\rmn{S}}$ magnitudes for which completeness is 50\% along the contour.}
 \label{fig:khalf}
\end{figure}

\section{The Data}
\label{sec:data}
The data set used, the reduction process and the photometric analysis and calibration have been extensively described in Paper I; hence we will only provide a short summary here.
NTT/SofI $J \, \rmn{and} \, K_{\rmn{S}}$ broad band observations of Wd~1 (centered on RA(2000) = $16^{\rmn{h}}47^{\rmn{m}}03^{\rmn{s}}$, Dec(2000) = $-45\degr 50\arcmin 37\arcsec$) and of a nearby comparison field (offset by $\approx 7\arcmin$ to the East and $\approx 13\arcmin$ to the South of Wd~1), each covering an area of $4\farcm5 \times 4\farcm5$ were retrieved from the ESO archive (PI: J.Alves).

Data reduction was performed using the eclipse jitter routines \citep{Devillard:2001fk}. Point Spread Function (PSF) fitting photometry was derived using the IRAF implementation of DAOPHOT \citep{Stetson:1987qy}. The number of objects detected in both $J \, \rmn{and} \, K_{\rmn{S}}$ bands is $\approx 7000$ for the Wd~1 field and $\approx 5300$ for the comparison field. Photometric zero points and colour terms were computed by comparison of instrumental magnitudes of relatively isolated, bright sources with counterparts in the 2MASS Point Source Catalogue \citep{Skrutskie:2006uq}. 

\section{2D Completeness maps}
\label{sec:2dmaps}

To obtain a correct cluster IMF and for the analysis of Wd~1 density profile, it is necessary to derive appropriate incompleteness corrections.
In Paper I the authors considered completeness correction as a function of magnitude and distance from the centre of Wd~1. In the present work we drop the assumption of radial symmetry and build  incompleteness correction maps as a function of the position on the chip and of the magnitude.

The main source of incompleteness in our case is crowding, which severely affects seeing limited observations \citep[see][as an example]{Eisenhauer:1998lr}. The effects of crowding on the detection of point sources change according to two quantities: the average stellar density and the magnitude contrast  between the given point source and its neighbours. Both of these quantities may not follow a radially symmetric or regular distribution. Very bright objects are normally scattered over the field in a nonuniform way. Even when they have a regular distribution, they still can cause sudden and very well localized drops in the completeness. In addition each of them has its own brightness and causes a lack of detections in areas of different angular width over the chip.
Stellar density itself does not \emph{a priori} have to follow a symmetric distribution; indeed the actual number of stars for a given position is determined  by an interplay of several factors, e.g. the intrinsic spatial distribution of stars within the cluster, varying extinction pattern (in the foreground, but also within the cluster) or changes in the fore- and background population characteristics, for example within the spiral arms.
For these reasons we think that an approach that does not assume any spatial distribution in the completeness characteristics of an observed field is preferable, in contrast to integrated or averaged cluster characteristics, and is definitely recommendable when spatial properties have to be investigated. For each photometric band we built a function with 3 variables:
\[
\mathcal{C}_j \equiv \mathcal{C} ( M_j | x, y, \umu)  \quad ;
\]
where $\umu$ is the actual value of the magnitude (in the $M_j$ band) and $(x,y)$ the position at which completeness is evaluated.
It is then possible to associate an incompleteness correction to each star for each photometric band.
The total incompleteness correction for a star detected in both $J \, \rmn{and} \, K_{\rmn{S}}$ bands is the product of the single corrections in each band. The reason is that each of these corrections represents the probability of detecting that given star in that specific band, and detections in each band are independent from each other.
The $\mathcal{C}_j$ completeness maps have been obtained in several steps, which are detailed in Appendix \ref{app:compmaps}.

A visualization of the completeness pattern for Wd~1 is shown in Fig. \ref{fig:khalf}. We display the $K_{\rmn{S}}$ band image of the cluster with superimposed 50\% completeness magnitude-loci. The contours are labeled with the corresponding values of $K_{\rmn{S}}$ magnitudes for which completeness drops to 50 \%.
Such contours follow the general distribution of stars, but also show peaks around the brightest stars, as expected; from  Fig. \ref{fig:khalf} it is clear that radial symmetry is not a perfect assumption for the completeness distribution of Wd~1.

\begin{figure}
 \centering
\resizebox{\hsize}{!}{\includegraphics{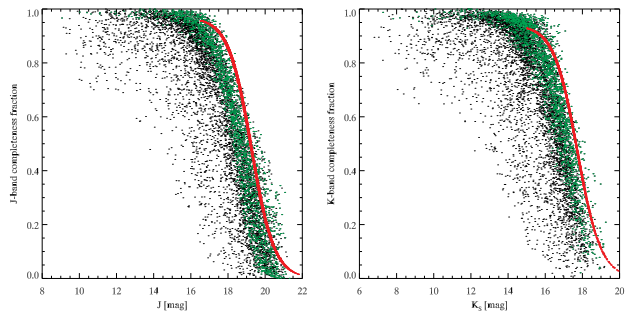}}
 \caption{Comparison of the completeness values for the Wd~1 stars (black dots) and the stars in the off-cluster frame (red dots)  for $J \, \rmn{and} \,  K_{\rmn{S}}$ bands. The green dots represent stars in Wd~1 frame, with angular distance from the cluster's centre larger than $2 \arcmin$.}
 \label{fig:compcomp}
\end{figure}

A comparison of the completeness values between Wd~1 frame and the off-cluster frame, for both photometric bands, is shown in Fig. \ref{fig:compcomp}. Given the spatial dependence of the completeness for the Wd~1 frame's stars, for them there is not an unique value of the  completeness at a given magnitude; for what concerns the off-cluster frame, we assumed spatial uniformity for the completeness, so the off-cluster frame stars (red dots) have unique values of the completeness as a function of magnitude (see Sect. \ref{subsec:compfield}). Figure \ref{fig:compcomp} shows that the completeness for the  off-frame stars is always higher, at a given magnitude, than the average completeness for the Wd~1 frame stars. Similarly, 50\% incompleteness is reached for the control field at $\approx 1 \,\rmn{mag}$ fainter than the average 50\% incompleteness for Wd~1 field. The cause of this difference may be found in the different degree of crowding of the two fields.
The green dots in the figure represent stars in Wd~1 frame located at more than $2 \arcmin$ from the centre of the cluster, corresponding to $\sim 2.3$ pc at the cluster's distance of 4 kpc (see Sect. \ref{subsec:distage}).
Even though these latter stars show - as expected - the highest completeness values for Wd~1 frame, they still have slightly lower completeness than the off-cluster frame stars. This is a reason to believe that crowding in this ''peripheral'' regions of the cluster frame is still higher than in the off-cluster frame, a hint to the presence of a low-mass cluster stellar population extending quite far away from the cluster centre. In Sect. \ref{sec:morph} we will show evidence that the low-mass stars of Wd~1 may indeed occupy a region with a radius of the order or even larger than 3 pc.

\section{Photometric errors}
\label{sec:everr}
\begin{figure}
 \centering
%\resizebox{\hsize}{!}{\includegraphics{Newerr_VS_mag.eps}}
\resizebox{\hsize}{!}{\includegraphics{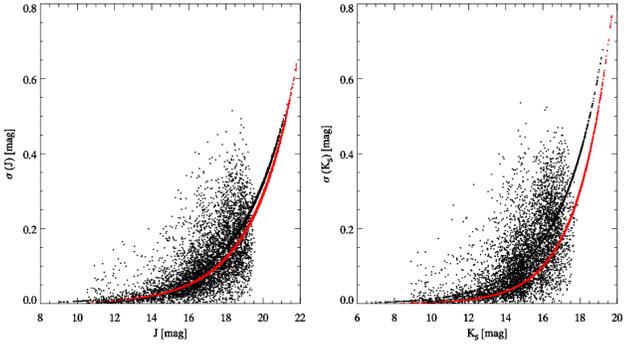}}
 \caption{The newly derived photometric errors as a function of magnitude in $J \, \rmn{and}\, K_{\rmn{S}}$. Errors for the stars in the cluster's field in black, errors for the off-frame stars in red.}
 \label{fig:newerrvsmag}
\end{figure}

%In this Section we will describe the methods we use to evaluate the photometric errors in the $J \, \rmn{and} \, K_{\rmn{S}}$ bands, as well as their correlation.
%A realistic estimate of these errors is crucial for our field-subtraction technique (see Sect. \ref{sec:subfield}), for the derivation of the global cluster properties (see Sect. \ref{sec:fundpar}) and for estimating  the uncertainties on the stellar masses derived by comparison with evolutionary models (see Sect.\ref{sec:IMF}).
As shown in Paper I, the DAOPHOT photometric errors are usually an underestimate of the true errors. DAOPHOT errors are connected to the residuals in the PSF fitting of the stellar counts. In principle this error estimate is absolutely correct for isolated stars only, so that the light is coming from the source of interest alone and only if the analytical PSF model chosen for PSF fitting is the correct representation of the true PSF shape. In this ideal case the errors would come only from the Poisson noise in stellar counts. In crowded fields, however, there are additional sources of uncertainty. The main one is the presence of bright objects. Even though the light from these sources is iteratively subtracted from the frame by the PSF fitting algorithm, the unsubtracted noise in the wings of these objects can still affect the magnitude estimate of nearby faint stars. Stellar crowding itself can cause problems when the algorithm has to disentangle very close sources even when they have similar magnitudes.
%In principle a bright background (e.g. from excited gas) with high spatial gradient can cause problems in DAOPHOT PSF fitting, but this is not an issue in Wd~1, where most of the gas has been removed by the many supernova explosions which already occurred during the lifetime of the cluster \citep[see][]{2006ApJ...650..203M}.
%For each simulated star, $i$ , and for each photometric band $M_j$, the difference between the input and output magnitudes, $\Delta M_j (i) = M_j^{out} (i) - M_j^{in} (i)$ were calculated. Then the stars were binned in magnitude bins and the average $< \Delta M_j >$ was used as the realistic error assigned to the detected stars in Wd~1 frame, according to their estimated magnitudes.
We use simulated stars to estimate realistic errors as a function of magnitude and position of the stars.
The new estimates of the  photometric errors are derived from the difference between the inserted and recovered magnitude of the simulated stars.
In addition, we examine the correlation between the estimated magnitude errors in the $J$ and the $K_{\rmn{S}}$ bands. 
The details of errors evaluation are given in Appendix \ref{app:everr}. Our error estimates are shown in Fig. \ref{fig:newerrvsmag} as a function of magnitude.

\section{Subtraction of the field stars}
We developed a technique for field subtraction based on a probabilistic approach. The technique takes into account the photometric errors, their correlation and the information about completeness.
%We consider the measurements in each photometric band as independent from each other and the photometric errors as Gaussians.
%This would not be true if the colour-magnitude diagrams had been used, which is often the case in the literature, being the colour errors strongly correlated with the magnitude errors (by definition); hence
%As shown in Sect. \ref{sec:everr} there is a correlation between $J$ and $K_{\rmn{S}}$ magnitudes; this correlation will be taken into account in the following. 
The natural space for our approach is an N-dimensional magnitude space. The technique is quite general and, as long as photometric errors in different bands and their correlations are evaluated, does not have to be limited to two bands. In the case of Wd~1 we only used $J$ and $K_{\rmn{S}}$ bands hence we will refer explicitly to them.

In the ideal case, a cluster magnitude-magnitude diagram (MMD) would look exactly the same as in the off-cluster field, plus additional stars belonging to the cluster, possibly following a separate sequence in the diagram, along an isochrone. It should be possible to compute the stellar densities at each MMD position for both the on-cluster and the off-cluster frames and compare them. Regions with an overdensity of stars would correspond to regions occupied by cluster members. %In the following, we will sometimes refer to the cluster field as the ON field and to the reference one as the OFF field.
The difficult part in the on-off density comparison  is to compute a proper density. Usually this is accomplished by gridding the CMD and by computing a density at each grid cell. Then, according to the numbers in the cluster cells and in the  off-field cells, some stars are subtracted, usually by making use of  Monte Carlo techniques. This approach has been very successful in many applications, also in our Paper I.
Anyway, any gridding or binning procedure always implies a loss of information.
Gridding is usually performed using equal cells, and this does not take into account, for example, the fact that photometric errors increase with magnitude, making it less obvious to which cell a faint star should belong. On the bright parts of the CMD the grid size may instead be very large compared to the photometric errors. In this case  the gridding would result in combining stars that, if errors would be reliable, are very distant from each other -- in units of their $\sigma_{\rmn{phot}}$ -- and then should not be considered ''similar'' and assigned to the same cell.
We decided to change this approach and to calculate the density of stars locally, at each position in the cluster's MMD where a star is located. Then we calculated the density in the same point of the MMD, but for the off-frame population. The ratio of the two densities is a measure of the membership probability of the star that is in that position in the cluster's MMD.

According to its photometric errors, each star is not a single point in the MMD, but a multi-dimensional Gaussian \emph{cloud} of probability, representing the chance of observing that object in that position. In our 2D case these Gaussians have an elliptical symmetry with semi-axis represented by $\sigma_J \, \rmn{and} \, \sigma_{K_{\rmn{S}}}$ and a tilt in the MMD related to the correlation between the two magnitude errors.
Since Gaussian probability is greater than 0 everywhere in the MMD, each star contributes \emph{a bit} to the total density at each MMD position, the closest stars to that position having higher weight.
Given a star with magnitudes $(J_*, K_{\rmn{S}\,*})$ we define the density at its position in the MMD in the following way:

\begin{eqnarray}
\label{eq:rhodef}
\lefteqn{\rho(J_*, K_{\rmn{S}\, *})  = \sum_i \,  \frac{1}{\mathcal{C}_{J_*} \, \mathcal{C}_{K_{\rmn{S}\,*}}} \times \frac{1}{\mathcal{C}_{J_i} \, \mathcal{C}_{K_{\rmn{S}\,i}}} \times \frac{1}{2 \upi |\mathbf\Sigma_*|^{1/2}} \times } \nonumber \\
\lefteqn{ \;  \times \frac{1}{2 \upi |\mathbf\Sigma_i|^{1/2}}  \times \int \exp \left[ -\frac{  (\mathbf{M} - \bmu_*)^T {\mathbf\Sigma_*}^{-1} (\mathbf{M} - \bmu_*) }{2} \right] \times } \nonumber \\
\lefteqn{ \; \times \exp \left[ -\frac{  (\mathbf{M} - \bmu_i)^T {\mathbf\Sigma_i}^{-1} (\mathbf{M} - \bmu_i) }{2} \right] \mathrm{d}\mathbf{M} \quad ;}
\end{eqnarray}
where the asterisk refers to the star at whose position the density is evaluated, such that $\mathcal{C}_{J_*} \, \rmn{and} \,  \mathcal{C}_{K_{\rmn{S}\,*}}$are the completeness fractions for that object, while $\mathcal{C}_{J_i}\, \rmn{and} \, \mathcal{C}_{K_{\rmn{S}\,i}}$are the completeness fractions for the other stars.
The density is calculated in both the on and the off field MMD, hence the index $i$ may run respectively on the stars in one or the other field.
The $\bmu$ vectors and the $\mathbf\Sigma$ matrix are, respectively, the measured magnitudes and  the covariance matrix associated to them:
\begin{equation}
\label{eq:defmusig}
\bmu_{*/i} = \left( \begin{array}{cc} J_{*/i} \\ K_{\rmn{S}\, {*/i}} \end{array} \right)  \quad  \mathbf\Sigma = \left( \begin{array}{cc}
\sigma^2_J                                  & r \sigma_J \sigma_{K_{\rmn{S}}}  \\
r \sigma_J \sigma_{K_{\rmn{S}}} & \sigma^2_{K_{\rmn{S}}}  \end{array} \right) \; .
\end{equation}
$|\mathbf\Sigma|$ is the determinant of the correlation matrix, 
and $r$ is the Pearson's correlation coefficient of equation (\ref{eq:Pearson}).
The $\mathbf{M}$ vector is the vector of coordinates $(J, K_{\rmn{S}})$ over which the integration is actually performed.
The integration is ideally performed in the whole (infinite) magnitude space. For obvious reasons we limit the numeric integration around each star to a region within $\pm 5 \, \sigma_*$ for each coordinate.

\label{sec:subfield}
\begin{figure*}
 \centering
%	\resizebox{\hsize}{!}{\includegraphics{Fieldsub_newerr.eps}}
	\resizebox{\hsize}{!}{\includegraphics{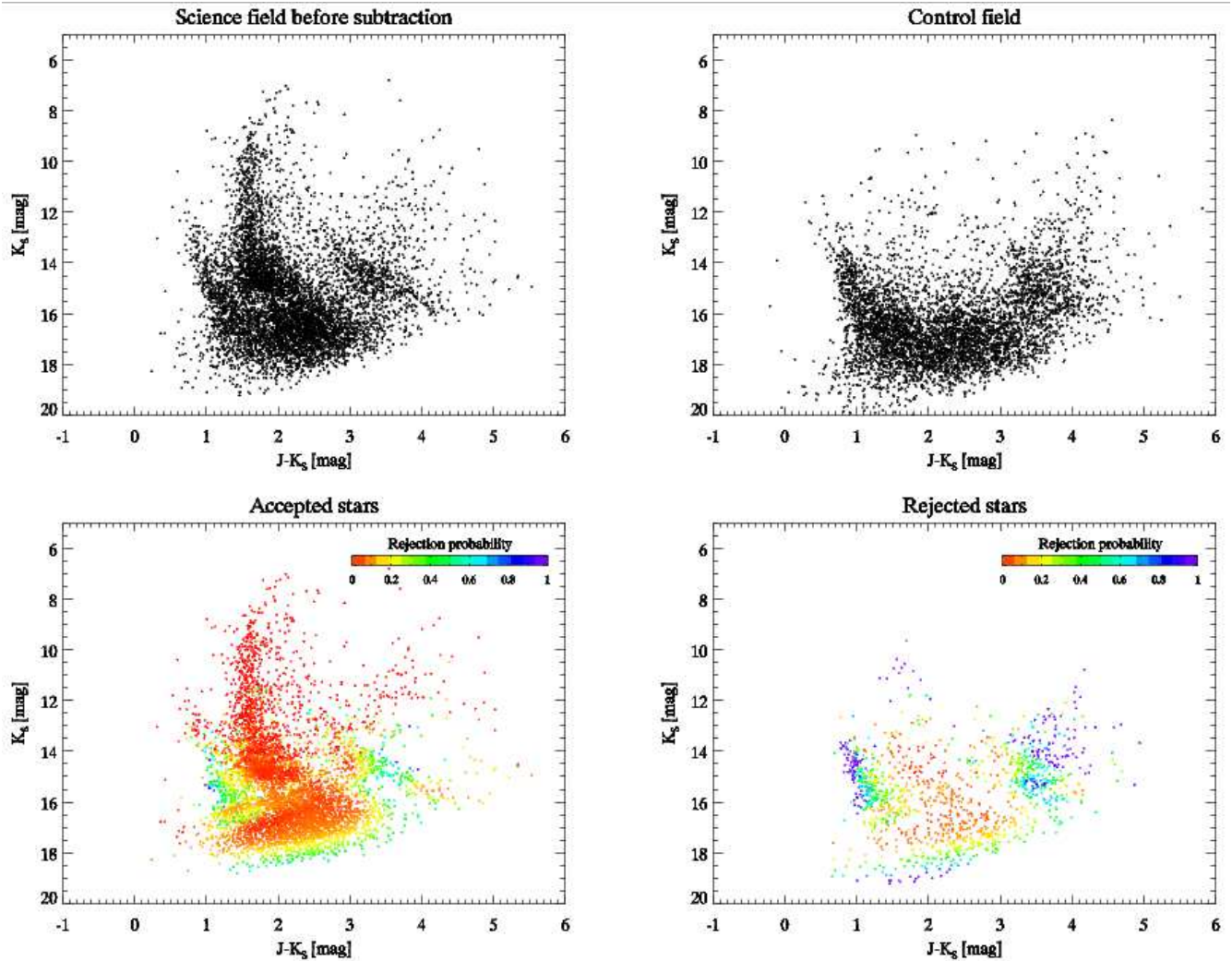}}
	\caption{ \emph{Upper panels}: on and off field CMDs. \emph{Lower panels}: Results of the subtraction process. The colour coding indicates the rejection probability $\mathcal{R}_{\rmn{rej}} = \rho^{\rmn{off}} / \rho^{\rmn{on}}$. Red represents very likely members ($\mathcal{R}_{\rmn{rej}} = 0$) and purple very unlikely ones ($\mathcal{R}_{\rmn{rej}} = 1$).}
 \label{fig:fieldsub}
\end{figure*}

Equation (\ref{eq:rhodef}) deserves several comments. Its meaning is the following: the contribution of the i-th star to the density at the $(J_*, K_{\rmn{S}\,*})$ position is the integral of the product of that star's probability distribution, convolved with the probability distribution of the *-th star. Then the total density in that position is the sum over all the $i$ stars either in the  on-field or in the off-field.
The probability of each single star is normalized to 1, as it has to be, but it is important to consider the completeness factors $\frac{1}{\mathcal{C}_{M_*}} \,\rmn{and} \,\frac{1}{\mathcal{C}_{M_i}}$ for $M = J, K_{\rmn{S}}$ that appear in equation (\ref{eq:rhodef}). These factors account for the missing detections in both the science and the control field. It is easy to understand why such correction is necessary.
Imagine a star in the cluster field, with completeness factor 0.25; it means that if we detected that object, then (in an statistical sense), there are 3 other similar objects that we were unable to detect. Now imagine that at the same position in the off-cluster MMD we would detect two objects both with completeness factors equal to 1. Neglect for a moment the real ''cloud'' shape of the stars' probability density distributions, and consider them \emph{ideally} as points in the MMD. By computing densities without  the completeness corrections, we would obtain $\rho^{\rmn{off}} =1\, \rmn{and} \,\rho^{\rmn{off}} = 2$. Hence we would \emph{oversubtract} that star from the cluster's MMD. On the other hand, the completeness factor tells us that the actual value of  $\rho^{\rmn{on}}$ is not 1 but 4, and then we would subtract that object only in 2/4 = 0.5 cases or, better said, we would assign to that star a 50\% membership probability (see also below).

Once we have both the on- and off-field densities at a given star's location in the MMD, we can compare them. The ratio $\mathcal{R}_{\rmn{rej}} = \rho^{\rmn{off}} / \rho^{\rmn{on}}$ defines a rejection probability; the higher the contrast in the two densities -- the lower $\mathcal{R}_{\rmn{rej}}$ is -- the more likely the object is a member. On the opposite side, when we are in a region of the MMD where no cluster members are present,  this number approaches 1.
Hence each detected object has its associated membership probability. To decide whether or not to keep it in the catalogue of member stars, we extracted uniform random numbers $\zeta \in [0,1]$. Then if $\zeta < \mathcal{R}_{\rmn{rej}}$ we discard the object, otherwise we keep it. 
This also means that in the following analysis the actual catalogues that we used may differ from one an other, because some stars may be sometimes excluded or included according to this random selection.
The uncertainties related to this selection directly propagate into, e.g., the IMF slope evaluation. To account for this we used multiple catalogue realizations and evaluate the uncertainties in the outcoming results as the scatter in the results (e.g. the IMF slope, see \ref{subsec:IMFdet} for more details).

In Fig. \ref{fig:fieldsub} we show the CMDs\footnote{Even though all the procedure is performed in the MMD, for the sake of clarity we show the most commonly used CMDs, where the usual characteristics of a cluster population are better visible} of Wd~1 frame and of the control frame, used as a reference for the field population, together with the results of the subtraction process. The colour coding in the lower panels indicates the rejection probability, $\mathcal{R}_{\rmn{rej}}$. As already mentioned in Paper I, and as is clearly visible in the upper panels of the figure, the foreground and -- especially -- the background population in the two frames do not look really similar. A possible cause for the differences might be a different amount of extinction along the different lines of sight in the on and off field. This population differences cause an under-subtraction of stars in certain regions of the CMD. However, it is clear that the most likely members in the lower left panel (the red points) follow a well defined cluster sequence; nevertheless some isolated fore and background stars in the cluster's frame also show an artificially high membership probability. The reason is that there are no objects in the off field MMD at the same position.

To avoid such artificial contamination, in addition to the subtraction process, we used a $\sigma$-clipping of our CMD (see Appendix \ref{subsec:sigmaclip}). After finding the best fitting isochrone (see Sects. \ref{sec:stellarmodels} and \ref{sec:fundpar}), we decided to keep  only stars that lie within 3 $\sigma$ from it, i.e. those stars that satisfy  the criterion $|J_* - J_{\rmn{isoc}}| < 3 \times \sigma(J_*) \, \rmn{and} \, |K_{\rmn{S}\,*} - K_{\rmn{S}\, \rmn{isoc}}| < 3 \times \sigma(K_{\rmn{S}\,*})$ for at least one point $(J_{\rmn{isoc}}, K_{\rmn{S}\,\rmn{isoc}})$ on the isochrone.

\section{The stellar models}
\label{sec:stellarmodels}
In the following analysis, we use a combination of Padova main sequence (MS) isochrones \citep{Marigo:2008lr} and Pisa pre-main sequence (PMS) models \citep{DeglInnocenti:2008fk}.
Padova models are accessible on the web \footnote{http://stev.oapd.inaf.it/cgi-bin/cmd} and are already provided in the 2MASS photometric system. For Pisa isochrones, we performed the conversion from the theoretical HR Diagram to the observational 2MASS-CMD ourselves. We used \cite{2005ESASP.576..565B} spectra, calculated with the PHOENIX model atmosphere code for the lowest temperature regions and \cite{2003IAUS..210P.A20C} spectra, based on ATLAS9 model atmospheres for the highest temperature in the PMS isochrones (see Table \ref{tab:trmassiso}). 
As in Paper I, we assumed a solar chemical composition for Wd~1, hence both the MS and PMS models used here have this composition. Nevertheless, given the intrinsic differences in the evolutionary codes (opacity tables, EOS, heavy elements mixture) and given also the fact that the ''Solar'' composition is not exactly the same in the two sets of models, they show some differences in the region of overlap.
Small differences are present also between the set of PMS isochrones transformed with PHOENIX and ATLAS9 model atmospheres.
We have carefully chosen the masses for the transition from one set of models to the other, in order to minimize the differences in colour between them.
The colour differences are shown in Table \ref{tab:trmassiso}, together with the mass and temperature ranges in which we adopt each model. The Pisa-ATLAS9 isochrones have been shifted in order to match the Padova isochrones at $4 \, M_{\sun}$, and the Pisa-PHOENIX have been shifted to match the Pisa-ATLAS9 ones at $2 \, M_{\sun}$. Table \ref{tab:trmassiso} shows that the offsets are quite small, specially when compared to the expected absolute precision in our photometry, which, taking into account the zero point errors 
 is of the order of 0.05-0.1 mag.

\setcounter{table}{0}
\begin{table}
 %\begin{minipage}{126mm}
\begin{center}
% use packages: array
 \caption{Mass, temperatures and magnitude offsets at the transition masses for our combined isochrones. The PS-AT9 offsets are: mag(Pisa-AT9) - mag(Padova) at $M = 4 M_{\sun}$ while the PS-PHX offsets are: mag(Pisa-PHX) - mag(Pisa-AT9) at $M = 2 M_{\sun}$.}
\label{tab:trmassiso}
\begin{tabular}{p{1.2cm}|p{1.3cm}p{2.1cm}p{0.5cm}p{0.5cm}p{0.5cm}}
\hline
Model 	& \centering Mass                & \centering T$_{\rmn{eff}}$                & \multicolumn{3}{c}{Offsets  [mag]}  \\ 
          	& \centering $[M_{\sun}]$    & \centering     [$10^3$K]    & $J$& $H$ & $K_{\rmn{S}}$ \\
\hline
Padova          & \centering  $M \geq 4 $           &\centering  T$_{\rmn{eff}} \geq 15.4$                    & -        & -        & -         \\ 
PS-AT9        & \centering  $2\! \leq \! M \!  \leq \!4$      & \centering $5.8\! \leq \! \rmn{T}_{\rmn{eff}} \! \leq \! 15.4 $&0.04 & 0.02 & 0.04\\
%PS-PHX        &  $ M \leq2 $           & T$_{\rmn{eff}} \leq 0.58 $          & -0.013 & -0.017 & -0.012\\
PS-PHX        &  \centering $ M \leq2 $           & \centering T$_{\rmn{eff}} \leq 5.8 $          & 0.01 & 0.02 & 0.01\\
\end{tabular}
\end{center}
%\end{minipage}
\end{table}

\section{Fundamental parameters of Wd~1}
\label{sec:fundpar}
Before proceding with the spatially dependent analysis, we derived the global, average properties of Wd~1 using the combined isochrones described above.

\subsection{Reddening and extinction}
\label{subsec:reddex}
For high mass stars on the MS, the near-infrared part of the spectrum is very well approximated by the Rayleigh-Jeans tail of a black body with temperature $T_{\rmn{eff}}$. Then, for masses above $\sim 5 \,\rmn{M}_{\sun}$, given that the SED shape is almost unchanged, the near-infrared, $J - K_{\rmn{S}}$, colours stay constant (and around 0 mag). The upper main sequence (UMS) in such a CMD looks simply like a vertical line. Therefore it is possible to estimate the reddening towards Wd~1 by fitting the $J - K_{\rmn{S}}$ colour of the UMS.
To perform the fit, we used the stars for which $K_{\rmn{S}} < 13.5$ mag and $1.2 < J - K_{\rmn{S}} < 2.0\,$ mag and minimized the quantity: 
\[
\sum_j | (J - K_{\rmn{S}})_j - (J - K_{\rmn{S}})_{isoc}| 
 \]
where j runs over the selected stars and the isochrone colour is taken at the same $K_{\rmn{S}}$ of the j-th star.
The  $J - K_{\rmn{S}}$ colour selection reduces the contamination by stars clearly belonging to the foreground or background population.
Once the $J - K_{\rmn{S}}$ reddening has been estimated, extinction $A_{K_{\rmn{S}}}$ is computed using an extinction law. Since, by definition, $E_{JK_{\rmn{S}}} = A_J - A_{K_{\rmn{S}}}$ we have:
\[
 A_{K_{\rmn{S}}} = \frac{E_{JK_{\rmn{S}}}}{\frac{A_J}{A_{K_{\rmn{S}}}} - 1} \,
\]
The knowledge of the interstellar extinction law provides the missing $A_J / A_{K_{\rmn{S}}}$ ratio.
While in Paper I we adopted the widely used \cite{1985ApJ...288..618R} --hereafter RL85-- extinction law, in the present work we use the much more recent \cite{2006ApJ...638..839N} one --hereafter N06--. The authors make use of a large number of red-clump stars located in the galactic plane. These stars have intrinsically similar colours, hence the observed differences in colour are related to different amount of interstellar absorption. Red-clump stars describe a straight line in the $(H-K_{\rmn{S}}, J-H)$ diagram parallel to the reddening vector. Hence the slope of this line can be used to determine the $A_J:A_H:A_{K_{\rmn{S}}}$ selective absorption ratios.
 In addition to the largely improved statistics, as compared to the few sources available in RL85, the N06 selective absorption has the advantage of having being measured using a $K_{\rmn{S}}$ filter, while RL85 used $K$. Hence the former provides a result that is in the same photometric system as our data.
The $J$-to-$K_{\rmn{S}}$ selective absorption ratio in the N06 case is given by $A_J / A_{K_{\rmn{S}}} = 3.021$, slightly higher than the $A_J / A_K = 2.518$ from RL85.
We checked that the obtained $A_{K_{\rmn{S}}}$ value actually does not depend on the age of the adopted isochrone. Our best fitting isochrone of 4 Myr (see also \ref{subsec:distage}) provides a value of $A_{K_{\rmn{S}}} = 0.907 \,\rmn{mag}$; if isochrones in the range 3--8 Myr are used, the scatter in the inferred $A_{K_{\rmn{S}}}$ is less than $0.01 \,\rmn{mag}$.
To estimate the error on the extinction value, we followed this reasoning. The absolute scatter in $J - K_{\rmn{S}}$ colour of the UMS stars used for the reddening fitting described above is about 0.2 mag. This means that a reasonable estimate for the reddening fitting error is 0.1 mag. From this, and using the N06 reddening law coefficients, it follows that  the error on the inferred total extinction can be estimated as $\Delta A_{K_{\rmn{S}}} = 0.05 \,\rmn{mag}$.

Given this errors and the results of Paper I, with $A_{K_{\rmn{S}}} = 1.13 \pm 0.03 \, \rmn{mag}$, it may seem that our new findings are inconsistent with the previous ones. Nevertheless one always has to keep in mind two crucial sources of systematic uncertainty in the method used and that are not included in the error estimates above. One is of course the choice of the stellar models, which may differ from one another both in the theoretical HR diagram and also in the transformations used to convert temperatures and luminosities into colours and magnitudes. A difference of 0.05 mag in the intrinsic near-infrared colours of UMS stars is anything but unexpected. We compared the Padova models used in the present work with the Geneva models used in Paper I \citep[][]{2001A&A...366..538L}, using in both cases solar metallicity and an age of 4 Myr and 3.9 Myr, respectively.  We observed differences in $J-K_{\rmn{S}}$ intrinsic colour ranging from 0.03 to 0.1 mag, at a given magnitude, in the mass interval from 5 to 30 $M_{\sun}$, used for the reddening estimate.
The other source of systematic uncertainty, is the aforementioned choice of the reddening law. In Paper I we used the RL85 law, and given that the selective absorption ratios are quite different between RL85 and N06, this explains the difference in our previous and new results for the total extinction.

To compare our findings with those by other authors, our best $A_{K_{\rmn{S}}}$value cannot be directly converted into an $A_V$ using only the N06 law. This law  has indeed been obtained only from $J$-band redwards \citep[see also][for the extension of the N06 reddening law towards photometric bands redder than $K_{\rmn{S}}$]{Nishiyama:2009lr}. Hence we use a combination of  $A_J / A_{K_{\rmn{S}}} = 3.021$ ratio from N06 and the $A_V / A_J = 3.546$ ratio from RL85 to obtain $A_V = 9.7$ mag.
As already noticed in Paper I different authors report values of $A_V$ that vary in the range from 9.4 to $\sim 12.0$ mag, so our final value is included well within this range.

Recently \cite{Negueruela:2010fj} have observed the presence of differential reddening across Wd~1. They report a range of $\Delta E_{VI} \approx 1.4 \, \rmn{mag}$. This range can be converted into a range of $\Delta E_{JK_{\rmn{S}}} = 0.51 \, \rmn{mag}$ using again a combination of RL85 and N06 laws, matched at the $J$-band. The observed colour range for the UMS members in our data set is somewhat smaller than this and part of this spread is probably due also to photometric errors and undetected binarity. Hence differential reddening across the cluster cannot be excluded, but  \cite{Negueruela:2010fj} extinction spread has to be regarded as an upper limit.
\begin{figure}
 \centering
	\resizebox{\hsize}{!}{\includegraphics{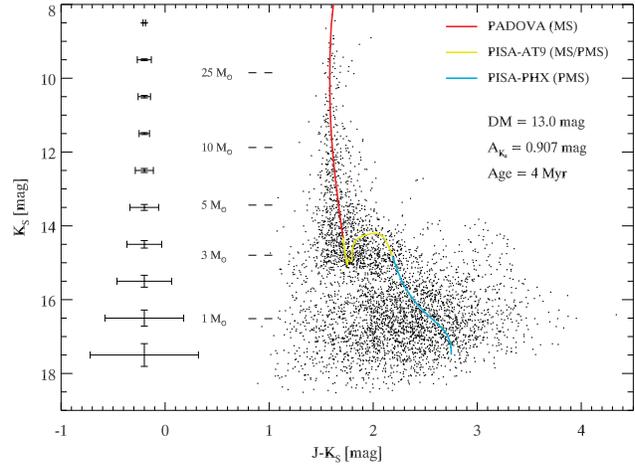}}
	\caption{One realization of the clean CMD after field subtraction and $\sigma$-clipping. Error bars represent the average photometric errors per magnitude bin. The best fitting combined isochrone is also shown, with three colours indicating the three different parts of which the isochrone is composed (see Sect. \ref{sec:stellarmodels} and Table \ref{tab:trmassiso}). Some values of the stellar mass are also shown for illustration.}
 \label{fig:CMDclean}
\end{figure}

\subsection{Distance and age}
\label{subsec:distage}

As illustrated in Paper I, the morphology of the PMS-MS transition region and of the whole PMS can be combined as a good age indicator for young clusters. Since extinction is determined independently (see \ref{subsec:reddex}), the distance modulus, DM, and the age, $\tau$, can be determined without having extinction as a free parameter.
Good age constraint is provided by those stars that have just entered the MS. These stars are located at the base of the vertical MS and have $14.9 \la K_{\rmn{S}} \la 15.1 \,\rmn{mag}$ and $1.6 \la J - K_{\rmn{S}} \la 1.8 \, \rmn{mag}$.
No cluster members are present at  magnitudes immediately fainter than that (see lower left panel in Fig. \ref{fig:fieldsub}). This Zero-Age-Main-Sequence (ZAMS) region is very well identifiable in the cluster's CMD and can be used to anchor the isochrones position.
It is worth mentioning that the determination of the age and DM in this paper is not done by a real fitting procedure, but through the conventional superposition of different isochrones for several values of the pair (DM,~$\tau$).
The DM and age values would be degenerate if only the ZAMS position would have been used for their determination. A slightly older isochrone would have an intrinsically fainter ZAMS point, and this could be compensated by a reduction of the DM.
Isochrones of different ages, however, also show different colours for the PMS branch, the younger, the redder.
Hence, in our comparison, after trying to reproduce the ZAMS point, we  also take into account the shape of the PMS-to-MS transition region and the PMS colour.
The uncertainty on the DM determination can be reasonably quantified as $\Delta \rmn{DM} =0.1 \,\rmn{mag}$ from the magnitude extension of the ZAMS region.
 The minimum age uncertainty that we can quote is instead half of the spacing between the different isochrones in our grid, i.e. 0.5 Myr.
By  isochrone superposition we obtain our fiducial values of $\rmn{DM} = 13.0\pm 0.1 \,\rmn{mag}$ (corresponding to a distance $d = 4.0 \pm 0.2 \,\rmn{kpc}$) and $\tau = 4 \pm 0.5$ Myr.

In Paper I we found values of DM = $12.75\pm0.10$ mag ($d = 3.55 \pm 0.17\,\rmn{kpc}$) and $\tau_{PMS} = 3.2$ Myr for the PMS population, while the MS stars provided weaker constraints on the age with $\tau_{MS}$ between 3 and 5 Myr.
The use of more recent PMS models partially reconciles our findings with those of other authors. For example \cite{Crowther:2006fk}, by comparing the number of WR stars and of cool hypergiants, find DM = 13.4 ($d = 4.8$ kpc) and $\tau = 4.5$ or 5 Myr. From observations of H\,{\sevensize I}, \cite{Kothes:2007fk} find a distance $d=3.9 \pm 0.7$ kpc.
\cite{Negueruela:2010fj}, from a comparison of their spectroscopically classified objects with models by \cite{Meynet:2000qy}, favour values of $d \ga 5$ kpc and $\tau \ga 5$ Myr. The authors point out the difficulties in spectral classification for several objects, the approximate character of the $T_{\rmn{eff}}$ scale, the uncertainties in $M_V$ values and, finally, the uncertainty in stellar evolutionary models for massive stars. The values of $d \sim 5$ kpc and $\tau \sim 5$ Myr are also supported by \cite{Ritchie:2010lr}, where the authors derive constraints on these quantities from the study a massive, interacting, eclipsing-binary.
Clearly there are still difficulties in the determination of the distance and age for Wd~1 with different methods providing slightly different values. Nevertheless with the present paper the differences between the values inferred using the intermediate- and low-mass end of the stellar population, on one side, and the high mass end on the other, are somehow reduced.

\section{The IMF of Wd 1}
\label{sec:IMF}

The comparison of observed magnitudes with isochrones allows the determination of stellar masses. %We used a new approach to compute the mass probability distribution for each star, taking into account the magnitude errors and their correlation. The distributions for the single stars are then combined to build the IMF of Wd~1.
The mass probability distribution for each star was determined by taking into account the magnitude errors and their correlation. The distributions for the single stars are then combined to build the IMF of Wd~1.
The detailed information on the completeness pattern across the field allows us to explore the variations of the IMF slope within Wd~1.
In the following, we consider all our objects as single stars, nevertheless we are aware of the possible biases introduced by neglecting the presence of binaries \cite[see][]{2009Ap&SS.324...95M}. This will be accounted for in an upcoming paper.

\subsection{The mass of the single stars}
\label{subsec:masssing}
Given our best-fitting isochrone (see Sect. \ref{sec:fundpar}) we used a maximum-likelihood approach to determine the mass of the member stars.
Again we work in the magnitude-magnitude space. There the probability density distribution of a star is characterized by its average magnitudes, by their photometric errors and by the correlation among them. Isochrones in the MMD are curves parameterized by the mass value of the star, $m$. Hence  the probability of a star with mass $m$ and magnitudes $\mathbf{M}(m) = (J(m), K_{\rmn{S}}(m)\,)$ to be observed at the $\bmu_* = (J_*, K_{\rmn{S}\,*})$ location in the MMD is:
\begin{eqnarray}
\label{eq:lkl}
\lefteqn{ p(m) = \frac{1}{2 \upi |\mathbf\Sigma_*|^{1/2}} \times } \nonumber \\
\lefteqn{\qquad \exp \left\{ -\frac{1}{2}  \left[\mathbf{M}(m) - \bmu_*\right]^T {\mathbf\Sigma_*}^{-1} \left[\mathbf{M}(m) - \bmu_*\right] \right\} }
\end{eqnarray}
(see equations (\ref{eq:rhodef}) and (\ref{eq:defmusig}) for a definition of the symbols).

Notice that $\int_{iso} p(m)\, \rmn{d}m = 1$, hence $p(m)$ represents a probability density distribution.
With this approach, we can determine not only the most likely mass for each star, by maximizing $p(m)$, but also the reliability of the mass  value obtained. If a star is indeed located very far from the best-fit isochrone (in units of its photometric $\sigma$), then its $p(m)$ will be a very \emph{broad} function, with a poorly determined peak. On the contrary if the star lies exactly on the isochrone then, ideally, $p(m)$ will be a Dirac $\delta$ function.

\subsection{IMF slope and total mass determination}
\label{subsec:IMFdet}

\begin{figure}
 \centering
	\resizebox{\hsize}{!}{\includegraphics{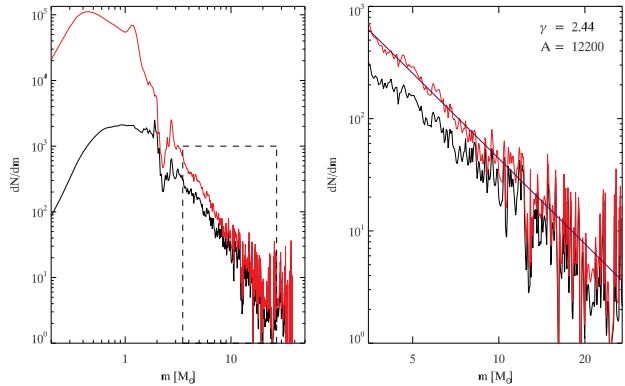}}
	\caption{\emph{Left:} Mass function for  Wd~1; the dashed box indicates the region that is used for the fit of the IMF slope. \emph{Right:} Zoomed version for the dashed box region; $\gamma =2.44 \, \rmn{and} \, \mathcal{A}=12200$ are our best estimates of the power law coefficient and IMF normalization constant, respectively. Red lines correspond to the completeness-corrected function while the uncorrected function is shown in black for comparison. The blue line in the right panel is the best-fitting power-law.}
 \label{fig:IMFglobal}
\end{figure}

A standard approach to evaluate the IMF slope of a cluster is to build a histogram of the stellar masses and then fit a power law (or a log-normal distribution) to the histogram. It is known, however, that the value of the slope is quite sensitive to the way the binning is performed and even to the space in which the fitting is done, i.e. a linear or logarithmic space for the mass coordinate \citep[see e.g.][for an exhaustive description of the subject]{2009Ap&SS.324...95M}.
These problems were also discussed in Paper I where we showed that the cumulative mass distribution, not requiring any binning, can be used to give stronger constraints on the IMF slope.
Here we introduce an alternative method that does not require any binning and makes use of the \emph{fundamental} information on the mass probability distribution, $p(m)$, which is always ignored when only the best-mass values are used, even without binning.
Given the $p_i(m)$ for each star in Wd~1 we define the observed Mass Function:
\begin{equation}
\label{eq:IMFdef}
\frac{\rmn{d}N(m)}{\rmn{d}m} = \sum_i \frac{1}{\mathcal{C}_{J_i}} \times \frac{1}{\mathcal{C}_{K_{\rmn{S}\,i}}} \times p_i(m)
\end{equation}

The $\frac{\rmn{d}N(m)}{\rmn{d}m}$ function for the whole Wd~1 population is shown in Fig. \ref{fig:IMFglobal}.
We used a restricted range of masses to determine the global slope of the IMF, $\gamma$, where $\frac{\rmn{d}N(m)}{\rmn{d}m} = \mathcal{A} \times m^{ - \gamma}$ with normalization constant $\mathcal{A}$ and $\gamma = 2.3$ for a typical Salpeter or Kroupa IMF, in the mass regime above $0.5 \,M_{\sun}$ \citep{Salpeter:1955lr,2001MNRAS.322..231K}.
The lower mass limit for the slope fit is chosen to be $m_{\rmn{min}} = 3.5 \,M_{\sun}$. At this mass we have 50\% global completeness on the whole frame. Locally this value could be different. For example, in the very centre of the cluster, high incompleteness is reached at high values of the stellar mass (see the lower panels of Fig. \ref{fig:IMFCompare}). This may cause some additional uncertainty on the derived IMF slope. 
The effects of spatially-varying-incompleteness are investigated in detail in Sect. \ref{subsec:spvarIMF}, where the potential bias in the cluster centre is also analyzed.
The upper mass limit for the slope fit is chosen to be $m_{\rmn{max}} = 27\, M_{\sun}$. The resons for this limit are: i)  the magnitude limit of our data-set. Stars more massive than this are above the linearity regime of the NTT/SofI observations that we have used. ii) stars above this mass are close to the turn-off region, according to Padova isochrones. Hence the determination of their initial masses starts to be age-dependent, and the complex post-MS evolution of such massive stars is quite uncertain, from the theoretical point of view. iii) the fitting procedure: above this mass value the numbers become so small that statistical fluctuation are not negligible and could lead to a bad fit.

We show the results in bi-logarithmic plots, but the actual fit has been performed in a linear space.
The global IMF slope we obtain is $\gamma = 2.44_{-0.08}^{+0.20}$, slightly steeper than an ordinary Salpeter/Kroupa IMF. We will explore in Sect. \ref{subsec:spvarIMF} local departures from this behaviour. For the normalization constant we found $\mathcal{A} = 1.22_{-0.14}^{+0.56} \times10^4$.
%The mean values and uncertainty of $\gamma \,\rmn{and}\, \mathcal{A}$  are evaluated by using 100 realizations of the catalogue of Wd 1 (see Sect. \ref{sec:subfield});  since there are slightly different members for each realization, the results of the IMF fit can be themselves different.
The best values and uncertainty of $\gamma \,\rmn{and}\, \mathcal{A}$  are evaluated by using a bootstrap technique, as detailed in Appendix \ref{subsec:IMFpar}. Given the couple of values $(\gamma_i, A_i)$ obtained from a single boostrap sample,  it is possible to associate to them a value of the total mass and total number of stars for Wd~1.
%Given the best-fit slope and normalization constant, we calculated the expected number of stars and the total mass of Wd~1. 
We extrapolate the power law with index $\gamma_i$ in the range $m[M_{\sun}] \in [0.5, 120] $. The upper mass limit is a reasonable estimate of the highest stellar mass that is expected to form in a massive cluster as Wd~1. From Padova isochrones, we have that stars with initial masses larger than $\sim 65\, M_{\sun}$ are supposed to have already undergone supernova explosions at the estimate cluster age of 4 Myr. Hence our results are estimates of the total initial mass and total initial number of stars for the cluster, under the assumption that the Present Day Mass Function is representative of the IMF value. For masses below $0.5 \,M_{\sun}$ and down to the hydrogen burning limit, i.e., $0.08 \,M_{\sun}$ we used the Kroupa IMF slope for this stellar regime, with a $\gamma = 1.3$.
In Appendix \ref{subsec:IMFpar} we also show how we derived the best estimates for the total number of stars and the total mass of the cluster, given the set of $N_{tot,i}$ and $M_{tot,i}$ from the different bootstrap samples.
The total number of stars is $N_{\rmn{tot}} = 1.04_{-1.83}^{+6.00}\times10^5$ while the total mass of the cluster is estimated to be $M_{\rmn{tot}} = 4.91_{-0.49}^{+1.79} \times 10^4 M_{\sun}$. %The relative errors in both $N_{tot} \, \rmn{and} \,M_{tot}$ are of the order of 10\%; these errors are estimated by varying $\mathcal{A} \, \rmn{and} \, \gamma$ within their uncertainty intervals.

Our present findings, based on a more complete and thorough approach, confirm the findings of Paper I and are on the lower end of the recent literature estimates for the mass of Wd~1. Using the MS turn-off mass and the identified post-MS member, by extrapolation of a Kroupa IMF down to lower masses, \cite{Clark:2005lr} found a somewhat higher value for the total mass of $\sim 10^5 M_{\sun}$.
Part of this discrepancy could be ascribed to the model-dependent uncertainties in the determination of the progenitor mass for the post-MS identified members. Additionally one has to be cautious when counting only the very massive stars to normalize the Kroupa IMF and then extrapolate it all the way down to low-mass stars. Only few young clusters in the Milky Way are known for which the IMF can be actually measured up to this masses, hence the nature of the IMF and its exact form is not known with great certainty in this regime. Moreover, also in the case that a standard IMF is valid for the very massive stars, high stochastic (Poissonian) fluctuations are expected when the numbers become small as towards the very high mass end of the Wd~1 population. 
A completely different approach was used by \cite{Mengel:2009uq} to determine a gravitating mass of Wd~1, $M_{dyn} =1.5_{-0.7}^{+0.9} \times 10^5 M_{\sun}$. The authors measured the radial velocity of $\sim 10$ stars from their spectra. From the dispersion of these velocity measurements,  the total mass of the system is derived, under the hypothesis of virial equilibrium, using the following equation:
$$
M_{dyn} = \frac{\eta \sigma^2 r_{hp}}{G} \, ;
$$
here $r_{hp}$ is the projected half-mass radius, $\sigma$ the velocity dispersion  and $\eta$ is a factor that the authors use under the additional assumption of isotropy.
Possible pulsations in the 5 yellow hypergiants (YHGs) of the sample, which would cause a wrong estimate of their radial velocities, may cause an overestimate of the true $\sigma$. \cite{Ritchie:2009lr} demonstrate indeed that one of the YHG observed in Wd~1, W243, shows a very complex, time-varying spectrum with signs of pulsation and mass loss that may hamper a precise determination of the radial velocity. This star is not in the \cite{Mengel:2009uq} sample, but it exemplifies that velocity dispersions derived from radial velocity measurements of evolved stars can lead to an overestimate of the true dispersion. In addition to this, we think that part of the discrepancy in the inferred dynamical mass could derive from the fact that Wd~1 is actually non spherical (see Sect. \ref{sec:morph}) and this anisotropy might be reflected also in the stellar motions. Therefore the $\eta$ factor used by the authors should be slightly modified, possibly giving better agreement with other findings. Indeed the velocity distribution seems to be not isotropic from our preliminary analysis of stellar proper motion using multi-epoch near-infrared AO data (Kudryavtseva et al. \emph{in preparation}).
On the other hand \cite{Fleck:2006qy} showed that the $\eta$ parameter is a time-dependent quantity, which changes rapidly, specially in very rich clusters, due to the effects of mass segregation. The authors found that the use of an $\eta \approx 10$, like in \cite{Mengel:2009uq}, may lead to \emph{underestimates} of clusters masses. We will show in the following that Wd~1 is mass-segregated. In such a case an increase in $\eta$ is needed to correctly estimate its dynamical mass. This would lead to an even stronger discrepancy with our photometric mass estimate. A possible interpretation of this difference could be that Wd~1 is indeed \emph{out of virial equilibrium}, with stellar motions still not relaxed after the gas-expulsion phase that followed the first supernovae explosions.
An effect that could balance the effects of mass segregation on the $\eta$ value is the inclusion of binaries in the estimates of this parameter. Binaries orbital motions increase the measured value of the velocity dispersion; consequently the true mass of a cluster is \emph{overestimated} if the binary contribution is not properly taken into account. \cite{Kouwenhoven:2008uq} showed the dependecy of the $\eta$ value on binary properties and cluster density. For the densest clusters ($N \ga 10^7 \, \rmn{stars}\, \rmn{pc}^{-3}$) the bulk motions dominate the total value of $\sigma^2$ while for the sparsest clusters, with $N \sim 0.1 \, \rmn{stars}\, \rmn{pc}^{-3}$, the velocity dispersion is fully dominated by orbital motions. Wd~1 density is in between these two extreme values. In this case the dynamical mass can be overestimated by 10-100\%, depending on the properties of the binary population. \citet*{Gieles:2010fj}, including mass-dependent mass-to-light ratio of stars and the intrinsically different binary properties of massive stars, found that the contribution to $\sigma^2$ from binaries orbital motions is already very important for young ($\sim10$ Myr), moderately massive ($M \sim 10^5 M_{\sun}$) and compact ( $r_{hp} \sim 1$ pc) star clusters, comparable to Wd~1.

\subsection{Spatial variability of the IMF}
\label{subsec:spvarIMF}

\begin{figure*}
 \centering
	\resizebox{0.29\hsize}{!}{\includegraphics{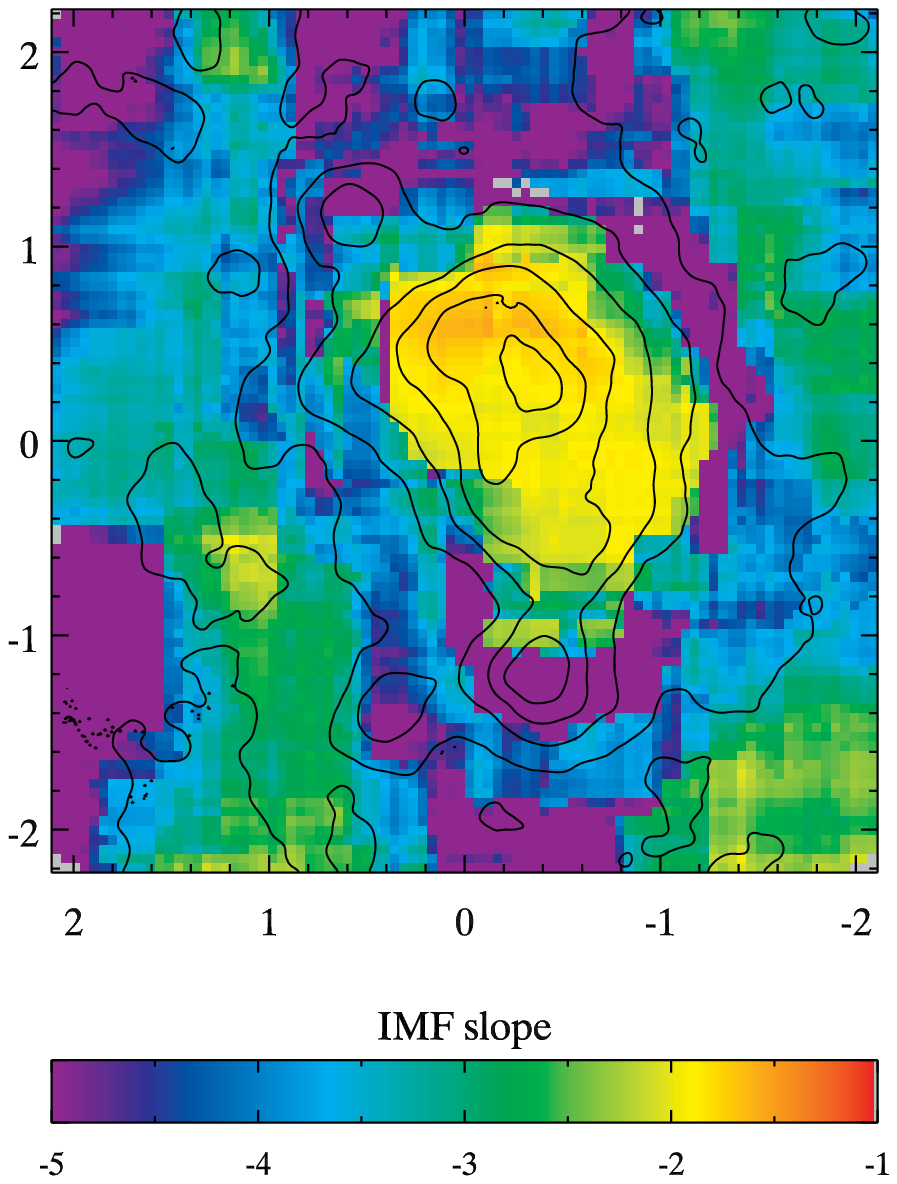}}
	\resizebox{0.29\hsize}{!}{\includegraphics{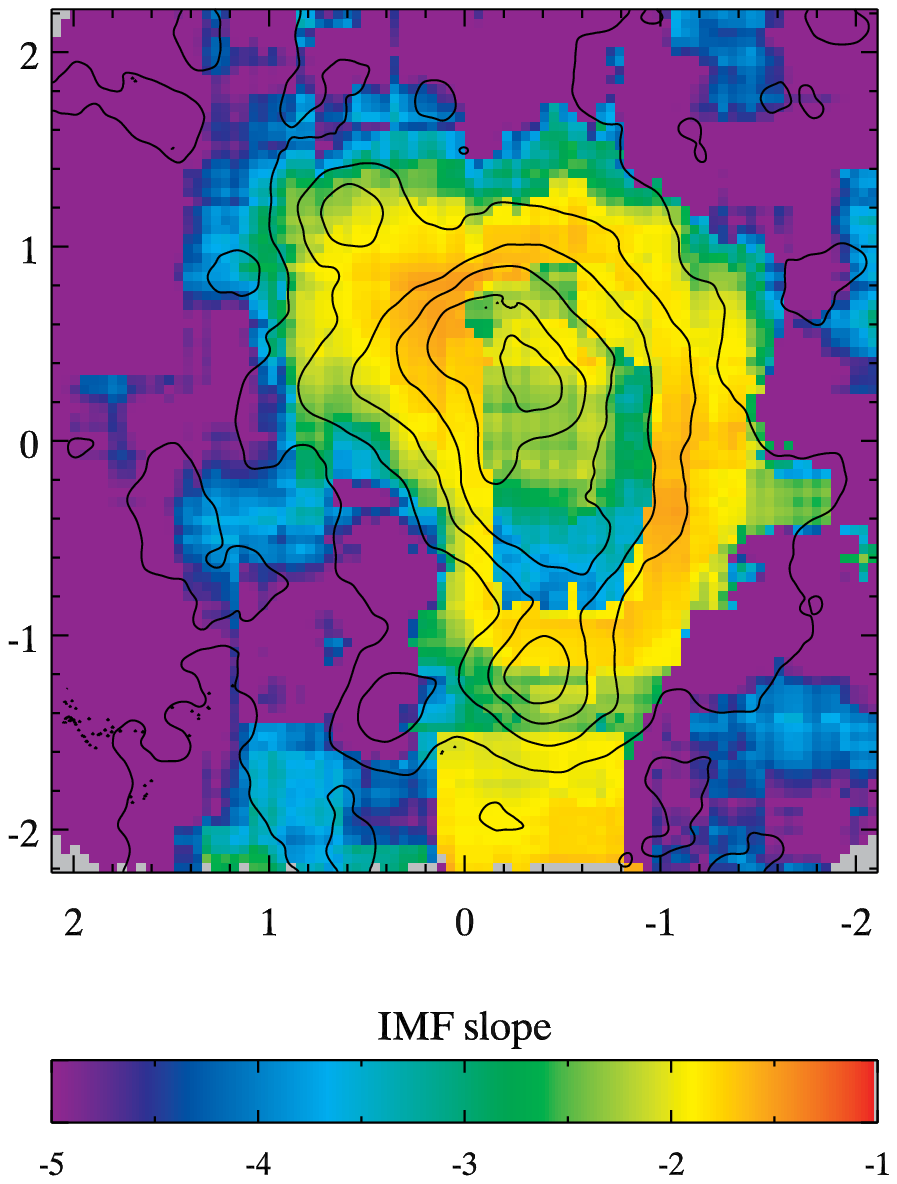}}
	\resizebox{0.29\hsize}{!}{\includegraphics{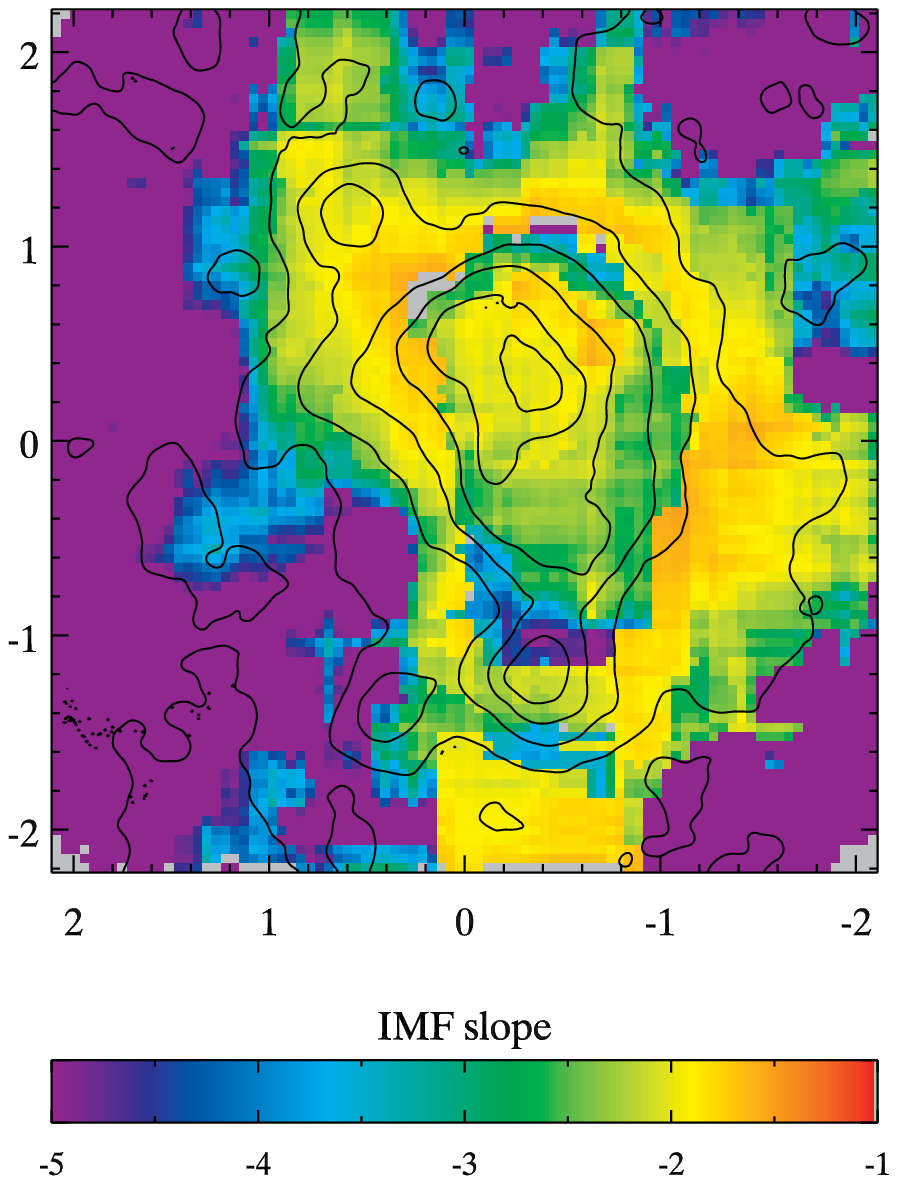}}
%	\resizebox{0.23\hsize}{!}{\includegraphics{IMFlocal_0p500_newerr_ALL.eps}}
	\caption{2D maps of the IMF slope of Wd~1. Within our definition, the colour coding corresponds to values of $-\gamma$. The three maps are built using only stars with a completeness factor, $\,\mathcal{C}_J \times \mathcal{C}_{K_{\rmn{S}}}$,  down to  0.125, 0.25, 0.375 (from left to right); overplotted are the contours of $J_{\rmn{half}} \times K_{\rmn{S}\, \rmn{half}}$. The x and y axes correspond to RA and Dec offsets, in arcminutes, relative to the centre of the reduced image, (RA, Dec) = $(16^{\rmn{h}}47^{\rmn{m}}06^{\rmn{s}}$, $-45\degr 50\arcmin 33\arcsec)$}
 \label{fig:IMFlocal}
\end{figure*}

\begin{figure*}
 \centering
	\resizebox{0.29\hsize}{!}{\includegraphics{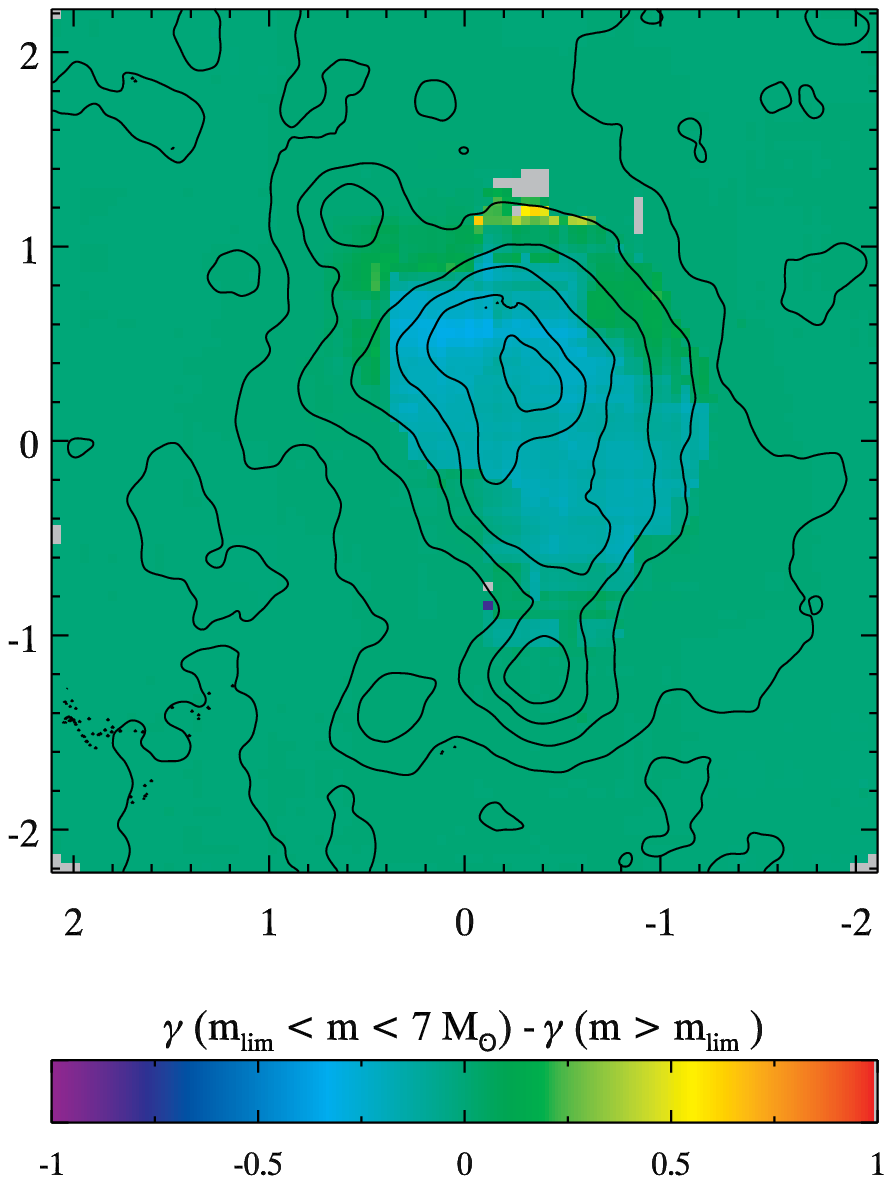}}
	\resizebox{0.29\hsize}{!}{\includegraphics{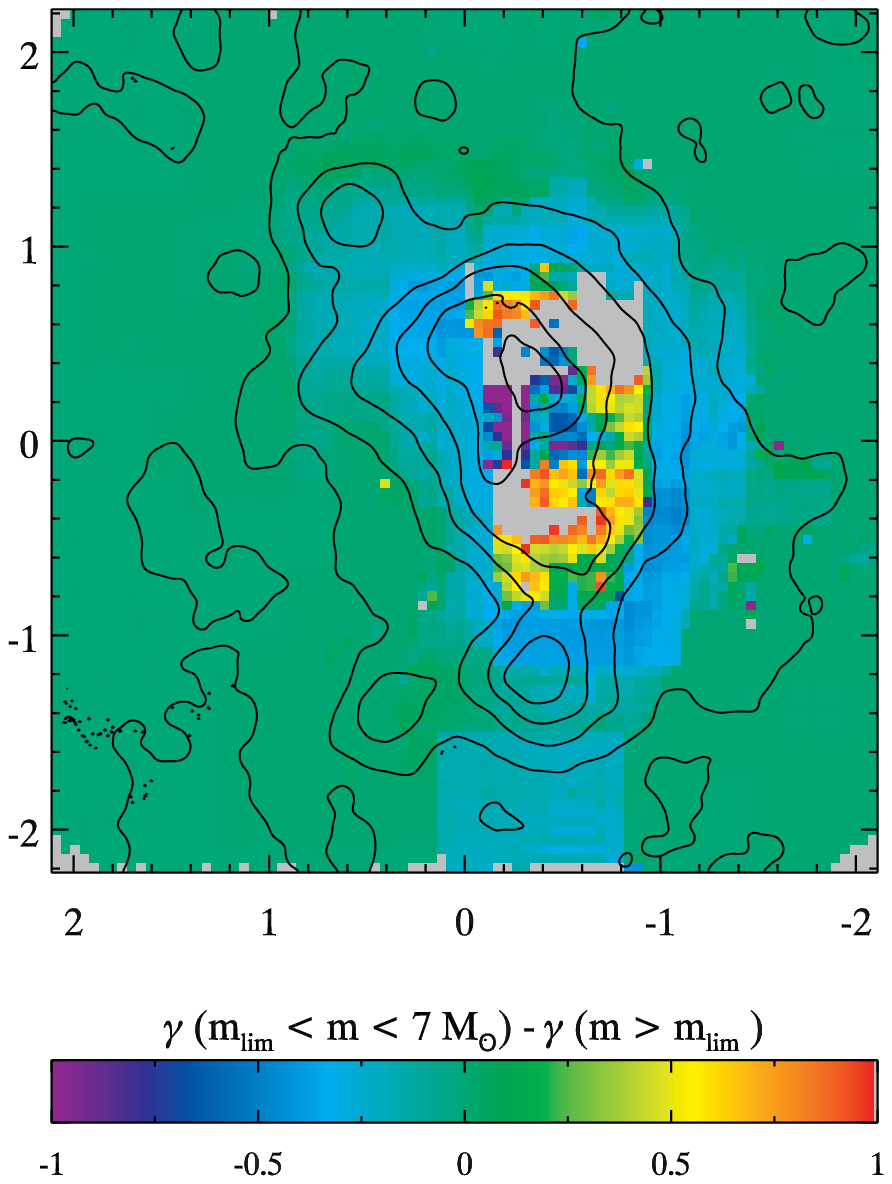}}
	\resizebox{0.29\hsize}{!}{\includegraphics{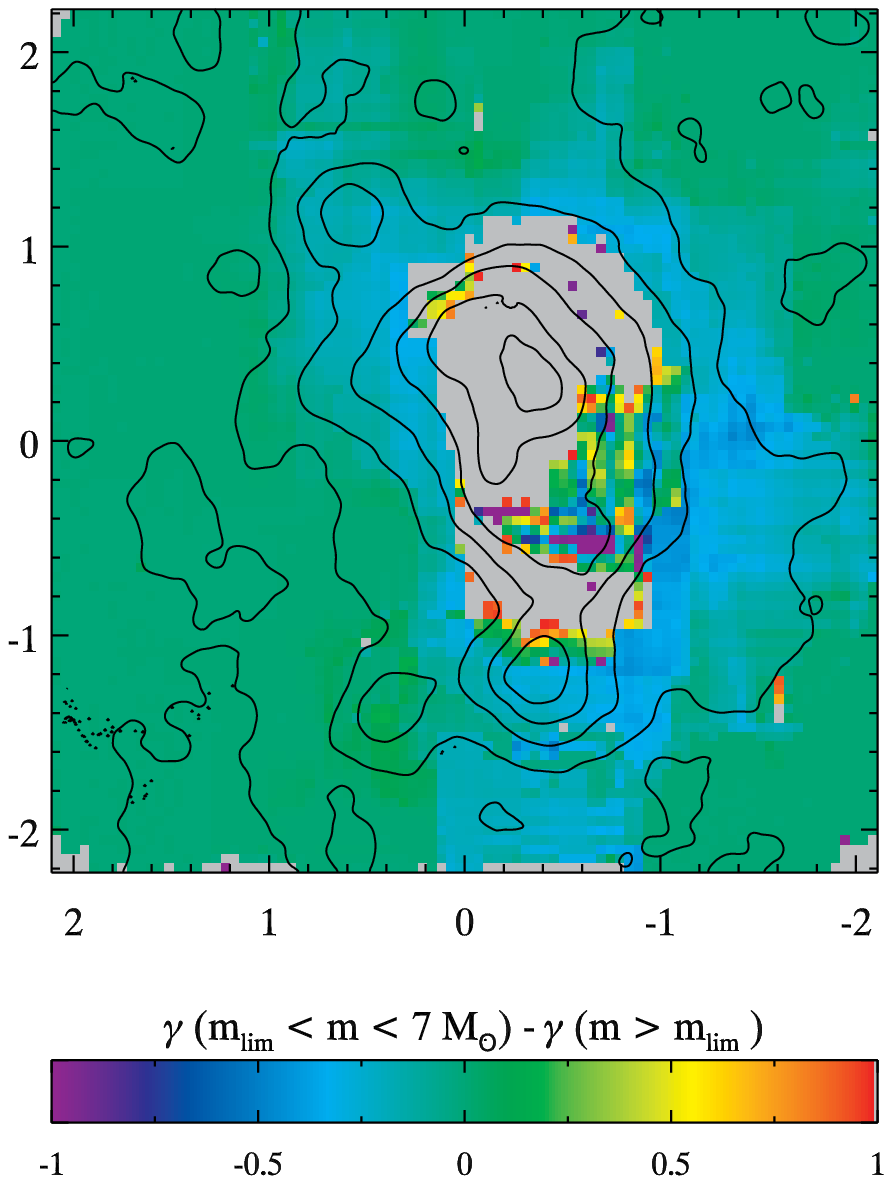}}
%	\resizebox{0.23\hsize}{!}{\includegraphics{IMFlocal_rat_d_0p500_newerr.eps}}
	\resizebox{0.29\hsize}{!}{\includegraphics{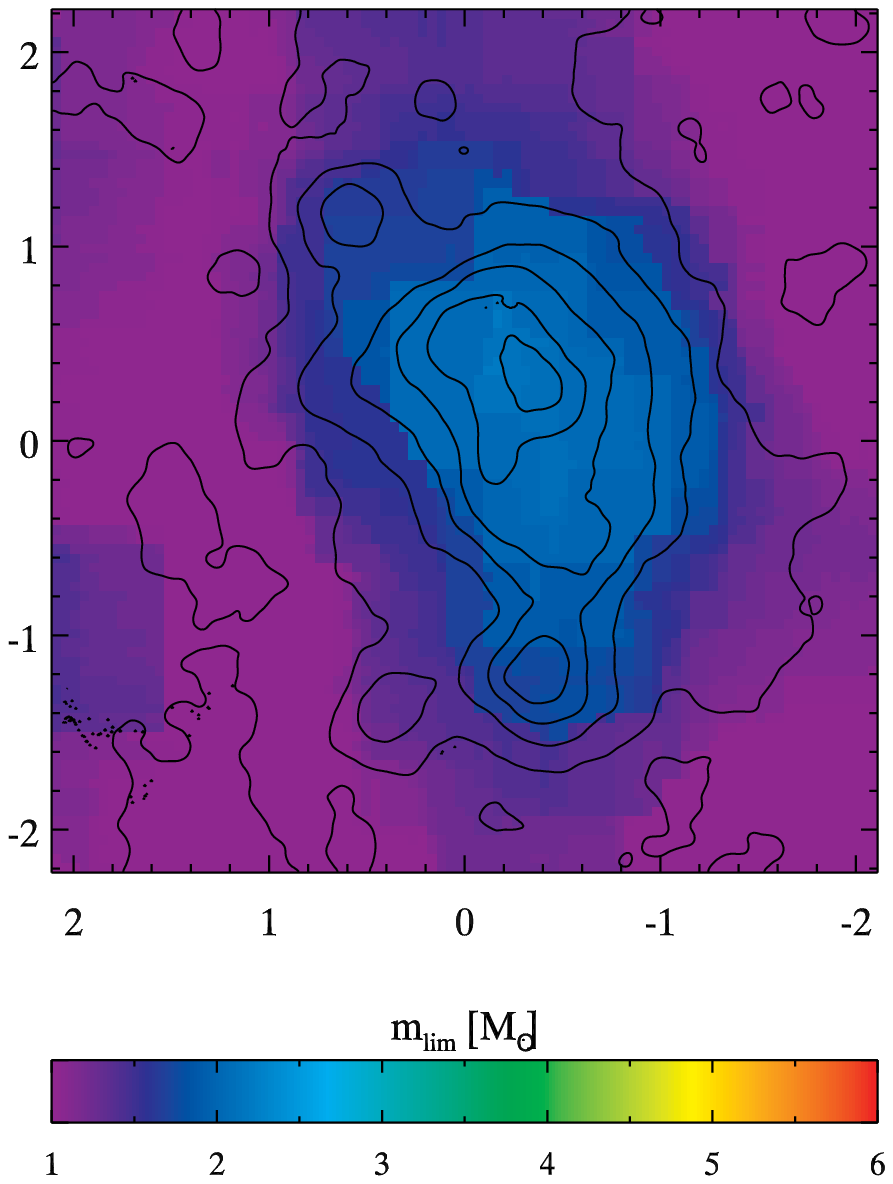}}
	\resizebox{0.29\hsize}{!}{\includegraphics{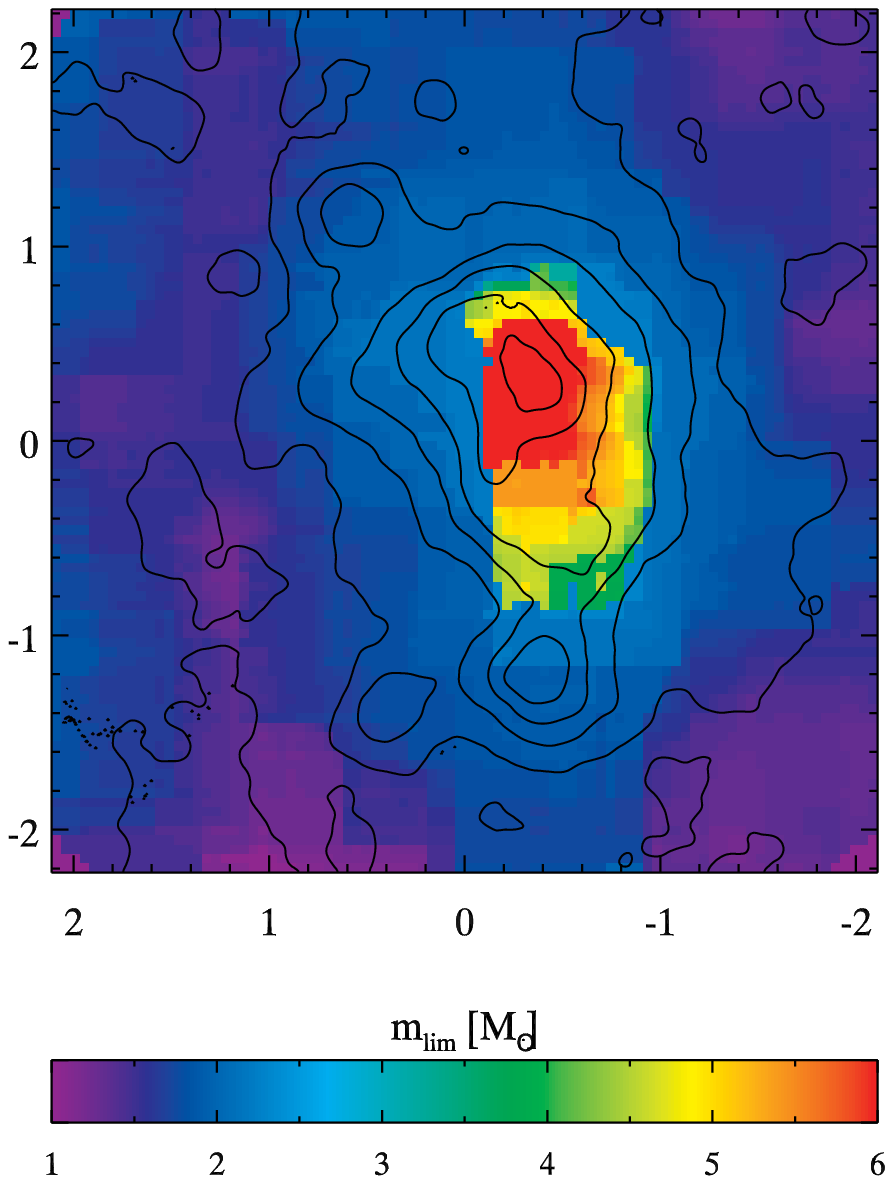}}
	\resizebox{0.29\hsize}{!}{\includegraphics{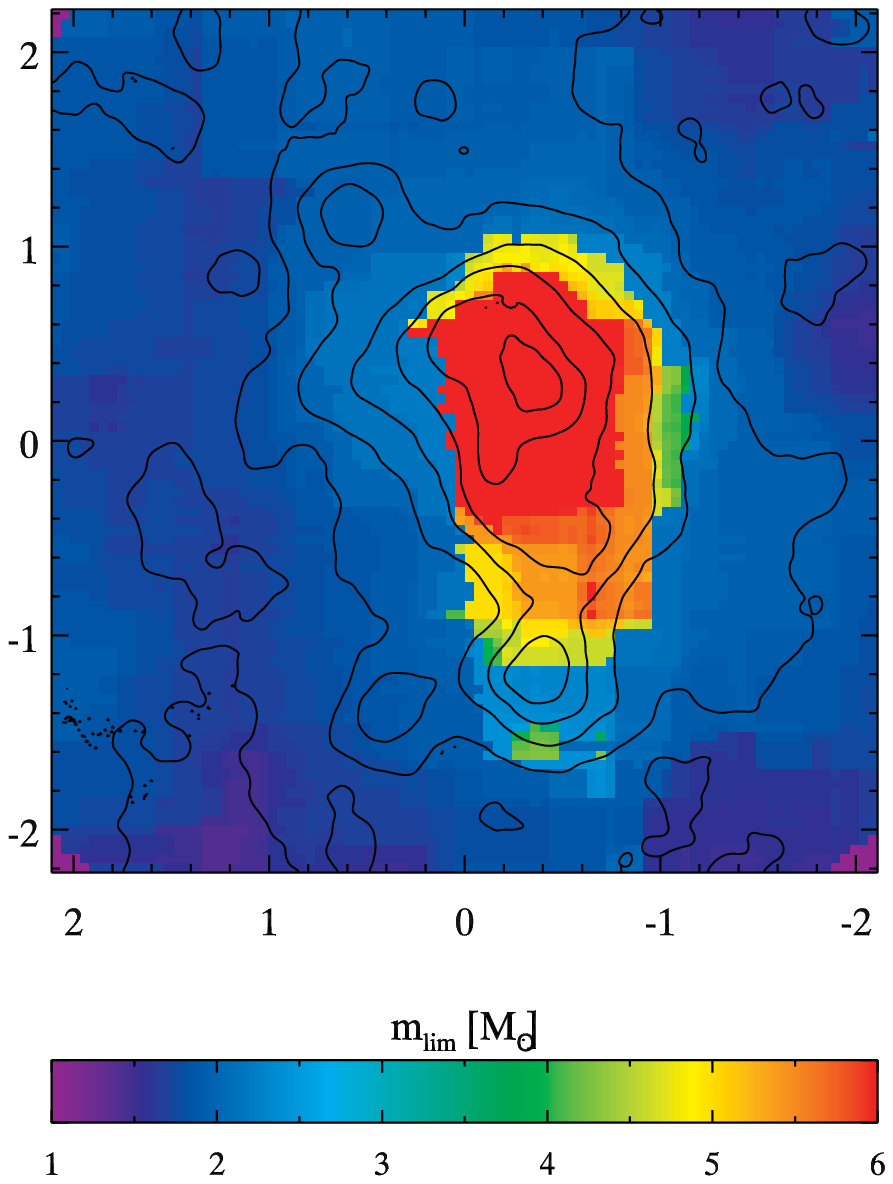}}
%	\resizebox{0.23\hsize}{!}{\includegraphics{IMFlocal_MASSLIM_0p500_newerr.eps}}
	\caption{\emph{Upper panels}: Differences in the $\gamma$ values using stars between the completeness-threshold mass and $7 M_{\sun}$ and using all stars above the threshold mass. \emph{Lower panels}: Values of the completeness-threshold mass. Columns from left to right correspond to $\mathcal{C}_J \times \mathcal{C}_{K_{\rmn{S}}} = 0.125, \, 0.25 \,\rmn{and}\, 0.5$ respectively. Gray areas are areas with not enough stars to perform a reliable fit of the IMF slope. The x and y axes meaning is the same as in Fig. \ref{fig:IMFlocal}}
 \label{fig:IMFCompare}
\end{figure*}

In Paper I we have shown that, considering concentric annuli centered on Wd~1 centre, and computing the slope of the IMF for the stars in the annuli, there is a tendency for a flattening of the IMF when going closer to the centre. The IMF slope was computed using stars more massive than  $3.4 M_{\sun}$ only, to avoid any bias due to the lower degree of completeness in the crowded centre of the cluster.
Still, close to the brightest stars, 50\% incompleteness is reached already at higher masses of up to $6 M_{\sun}$.
This may still cause an artificially flatter IMF in the central parts of Wd~1, because, even though the IMF slope is obtained using the incompleteness corrected number of detected stars, the correction itself becomes quite uncertain when one uses it for much lower levels than 50\% completeness.
We have also shown that the completeness pattern in Wd~1 is not really radially symmetric and we will show in the following Sect. \ref{sec:morph} that the shape of the cluster itself is elongated; hence using concentric annuli can smooth out some of the intrinsic spatial variations of the IMF.
With our new approach we determine the IMF slope locally, in  order to follow its real pattern within Wd~1.

To calculate the IMF slope at each position, we used a moving box, 200 pixels in size. The slope was obtained by the same technique described in Sect. \ref{subsec:IMFdet}, applied only to the stars in the box. At each position we additionally selected stars such that the total completeness factor $\mathcal{C}_J \times \mathcal{C}_{K_{\rmn{S}}}$ is always higher than some fixed threshold values. In this way we can compare results for varying completeness thresholds. Hence, at each position the minimum mass considered can be different. The upper mass limit is determined by the stochastic distribution of non-saturated high-mass stars within the moving box.
The fit is performed only when the number of stars inside the box is larger than 30.
The calculation is repeated at each pixel. Anyway we are forced to use a moving box that is much bigger than the sampling scale because we need enough stars to perform the IMF slope fit. Hence the adjacent-pixel slope-values are not  independent from each other. The final maps are obtained by convolving the adjacent-pixel slope-values with a Gaussian kernel of FWHM = 100 pixels (half-box size)
At the distance of Wd~1 ($\approx 4 \,\rmn{kpc}$) and with a plate scale of $0\farcs29 / \rmn{pixel} $ 100 pixels corresponds to $ \sim 0.5 \,\rmn{pc}$ in linear scale.

Results for three values of $\mathcal{C}_J \times \mathcal{C}_{K_{\rmn{S}}} =$ 0.125, 0.25 and 0.375, are shown in Fig. \ref{fig:IMFlocal}. With our definitions, the \emph{conventional} 50\% threshold in one band is replaced  by a $0.5 \times 0.5 = 0.25$ combined threshold.
The green areas correspond to an ordinary Kroupa-like slope ($\gamma \approx 2.3$). The yellow-red areas in the centre indicate regions with a flatter IMF, i.e., with more high-mass stars than what is predicted by a Kroupa IMF. Blue-purple areas are areas with a paucity of massive stars. Hence, when fitting a power law we obtain a very steep function due to the overabundance of low-mass stars. 
%Gray areas, barely visible in the corners of the plots, correspond to regions where the number of stars within the box is too low to perform a significative fit of the IMF slope; this is simply a boundary effect, due to the fact that the effective area within the frame, for a box centered on a location close to the boundaries, is smaller than $200 \times 200 \,\rmn{pixel}^2$ of a box located entirely inside the frame.
The contours in the three panels are the contours for the $\mathcal{C}_{J_{\rmn{half}}} \times \mathcal{C}_{K_{\rmn{S, half}}}$ function, i.e. the product of the $J \, \rmn{and} \, K_{\rmn{S}}$ 50\% completeness magnitudes. These contours trace the shape of the total completeness correction factor.

The overall pattern of the IMF slope remains unchanged among the three different maps of Fig. \ref{fig:IMFlocal}. Nevertheless, some differences can be noticed.
Going to lower and lower completeness thresholds (i.e., from right to left in the figure), the yellow region in the centre of the image ''shrinks'', leaving space for regions of slightly steeper IMF around it. 
Hence, when completeness corrections are properly taken into account, there are strong hints that low-mass stars are overabundant outside the very centre of Wd~1. On the other hand, the yellow-red regions still visible in the centre of the maps, even for the lowest completeness threshold, indicate an overabundance of massive stars that is very likely to be intrinsic and not just a result of missing detections in the low-mass end.
A similar ''shrinking'' behaviour is observed for the ''purple'' outer regions that are very pronounced in the two rightmost panel of the Fig. \ref{fig:IMFlocal} and less in the leftmost. In this case the effect is due to the difference in the mass intervals used for the fit of the IMF slope. At higher completeness thresholds, only few mass-points are available, and the differences in number counts between the low-mass and high mass limits within the fitting interval are very high when few high-mass stars are present. When lowering the completeness threshold, star counts are added at lower masses, hence the observed mass function becomes more ''regular'' and the results of the fit of the slope are less extreme.

Given the low number of stars towards the high-mass end, statistical fluctuations in this regime may increase the uncertainty of the IMF slope. To see whether this effect is important, we compared the $\gamma$ values for all stars with $m > m_{\rmn{lim}}$ and the $\gamma$ values for stars with $m_{\rmn{lim}} < m < 7 \,M_{\sun}$ only. The results are shown in the upper panels of Fig. \ref{fig:IMFCompare}. 
The grey regions in the centre are regions where the number of stars in the fitting interval was too low to perform a good fit.
Excluding these regions, it is clear that the difference between the two slopes is almost everywhere zero. This tells us that the fit is dominated by the low-mass regime of the fitting interval, where the stars are more numerous and where the overall shape of the IMF is very well determined since statistical fluctuations are less pronounced.
The only differences between the two slopes are observed in the very centre, where $m_{\rmn{lim}}$ becomes very close to the upper-mass limit  of $7 \,M_{\sun}$. In these regions, indicated by a cyan colour, the inferred IMF is flatter when the high-mass end is neglected.  Anyway, the incompleteness level in the very centre is high, hence these small differences ($\Delta\gamma \la 0.3$) cannot be considered significant.

%actually this is not the case. The IMF slope obtained considering only stars with $m_{\rmn{lim}} < m < 7 \,M_{\sun}$ looks flatter only because the completeness corrections in the very centre are less reliable. Thus it is not possible to account for the missing detections in the very centre, at low stellar mass, where the incompleteness severely affects already  $4-5\, M_{\sun}$ stars. 
%This result provides an other indication of mass segregation. Even though, it also hints to the need of observations with better angular resolution in the very centre of Wd~1. This would allow to observe stars to much lower masses than here, where we are forced to use data obtained in seeing limited conditions and then would allow a direct solution of the real/completeness-biased mass segregation controversy.
Summarizing, we can say that the overall IMF shape is consistent with a Salpeter or Kroupa galactic IMF in the range of masses between 3.5 and 27 $M_{\sun}$. This slope is the spatial average of a slope that varies across Wd~1. 
%A trend in the local IMF slope values is clear, with central regions having flatter IMF compared to the outskirts of the cluster.
A trend in the local IMF slope values can be observed in Fig. \ref{fig:IMFlocal}, with central regions having flatter IMF compared to the outer regions of the cluster. This is a robust indication that Wd~1 is mass segregated. We will show additional evidence of this mass segregation in Sect. \ref{sec:morph}, where we will also discuss its possible origins.

\section{Morphology of Wd~1}
\label{sec:morph}

\begin{figure*}
 \centering
	\resizebox{0.44\hsize}{!}{\includegraphics{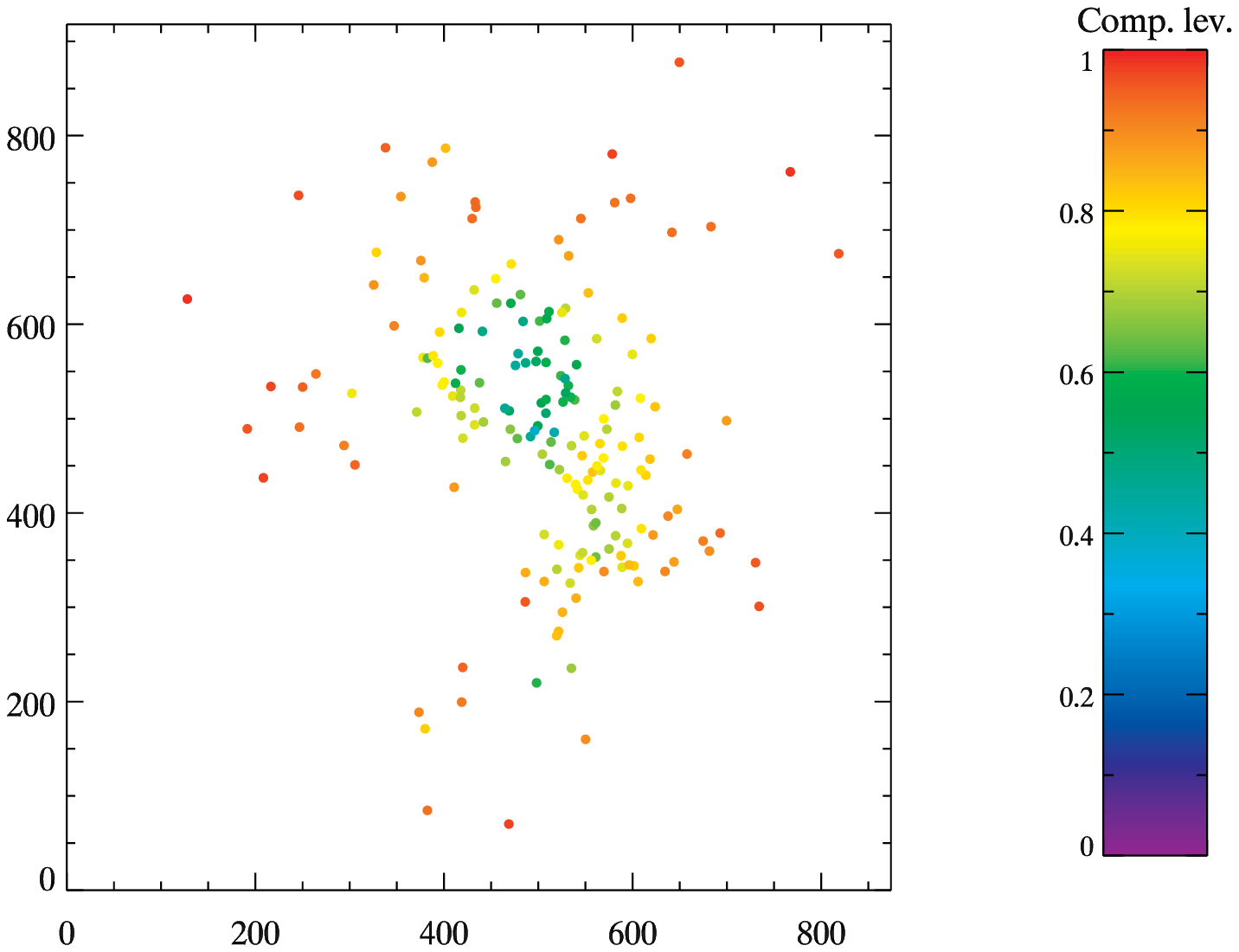}}
	\resizebox{0.44\hsize}{!}{\includegraphics{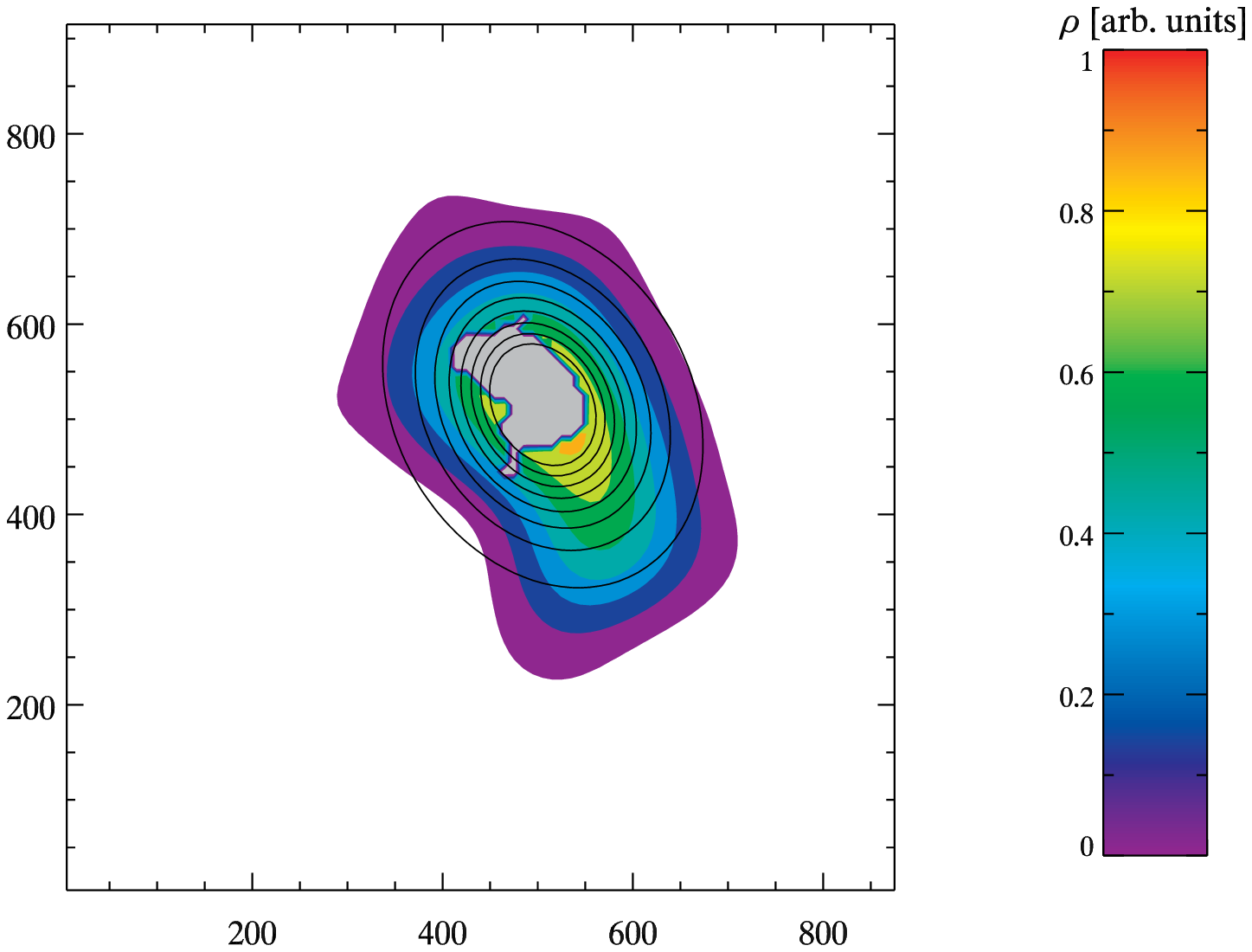}}
	\caption{Example of surface density contours and a fit using the GEFF profile, in pixel coordinates on the frame. \emph{Left}: stars with $m > 7.5 \,M_{\sun}$ and $\mathcal{C} > 0.375$; completeness factors for the single stars are colour-coded. \emph{Right}: the corresponding density contours (in colours). The density is in arbitrary units and the colour-coding goes from the minimum to the maximum density. The gray area corresponds to the area where the average completeness for the whole cluster's population is below 0.25; this area is masked-out when performing the GEFF profile fit. The results of the fit are displayed as black elliptical contours.}
 \label{fig:dencon}
\end{figure*}

Several recent studies indicate that Wd~1 is elongated \citep[see e.g.][]{2006ApJ...650..203M}. In our Paper I, assuming an elliptical shape with $a \,\rmn{and}\,b$ as semi-major and semi-minor axis respectively, we found an ellipticity of the cluster, $\eta = 1 - \frac{b}{a} = 0.19$, when stars with masses in the range 10 to 32 $M_{\sun}$ were considered. The value slightly decreased, to $\eta = 0.15$, using masses between 3.5 and 10 $M_{\sun}$. Elongation was computed by calculating the half-mass radius as a function of the position angle (PA), considering for each PA only stars within $\pm 45 \degr$ around PA and, correspondingly, around PA + $180\degr$.
%We also showed that the overall density profile of the cluster follows a $\rho(r) \propto \left[1 + (r/\alpha)^2\right]^{-\beta} $ radial law  \citep[see][hereafter EFF87]{Elson:1987lr}, with core radius related to the $\alpha$ parameter by EFF87 equation (22), i.e., $r_{\rmn{c}} \approx \alpha(2^{2/\beta} -1)^{1/2}$ and $\beta = 2$ for Wd~1.
We also showed that the overall surface mass density profile of the cluster follows a $\Sigma(r) \propto \left[1 + (r/\alpha)^2\right]^{-\beta} $ radial law  \citep[see][hereafter EFF87]{Elson:1987lr}, with core radius related to the $\alpha$ parameter by EFF87 equation (22), i.e., $r_{\rmn{c}} \approx \alpha(2^{2/\beta} -1)^{1/2}$ and $\beta = 2$ for Wd~1. At large distances from the center, the 3D density profile goes like $\rho(r) \propto r^{-2 \beta -1}$ -see EFF87 equations (13a) and (13b). Hence an index $\beta = 2 $ for the 2D density profile implies a 3D density that goes like $r^{-5}$, which corresponds to a \cite{Plummer:1911mr} model. A $\beta = 0.5$ corresponds instead to an isothermal sphere with 3D density going like $r^{-2}$. The density profile of Wd~1  falls more rapidly compared to the case of R136 cluster in the Large Magellanic Cloud. This cluster has a mass comparable to that of Wd~1 and slightly younger age of $\sim3$ Myr but shows a profile that is closer to isothermal, with $\beta \approx 0.8$ \citep{Andersen:2009xt,Campbell:2010hc}

%Our new approach for incompleteness correction provides the opportunity of a 2D study of the cluster's stellar density distribution.
Our 2D incompleteness mapping enables a study of the cluster's 2D stellar density distribution.
We calculated the surface number density for several mass ranges and used  four values for the completeness threshold.
Given a lower-mass threshold, $m_{\rmn{low}}$, and a completeness threshold, $\mathcal{C}_{\rmn{tr}}$, we considered all the stars above these thresholds for calculating the stellar surface number density. The number density was computed using a moving box 100 pixels in size\footnote{This is half of the box size used for the computation of the local variations of the IMF (see Sect. \ref{subsec:spvarIMF}). Since here we just need to count the stars, we do not need large numbers within the box.}, which was moved pixel-by-pixel. After counting the stars at each position, we convolved the counts with a Gaussian kernel of FWHM = 50 pixels, i.e. half-box-size, to account for the fact that the density value computed at each pixel position is not independent of the values computed at nearby pixels.
In this way we have been able to build smooth number density maps for Wd~1.
These density profiles are always elongated, hence we decided to perform a fit by using an elliptical generalization of the EFF87 profile, a natural extension of the work done in Paper I. We will refer to this profile as GEFF.

The GEFF profile can be described in the following way:
\begin{equation}
\label{eq:GEFF}
\Sigma_{\rmn{GEFF}} = \Sigma_{\rmn{BG}} + \Sigma_{\rmn{c}} \left(  1 + L^2  \right)^{-\Gamma} \quad ;
\end{equation}
where $\Sigma_{\rmn{BG}}$ is a stellar background density, $\Sigma_{\rmn{c}}$ is the density in the centre and $\Gamma$ represents the density decay for large distances from the centre $(L \gg 1)$.

The quantity $L^2$ is given by:
\begin{equation}
\label{eq:Lsquare}
L^2 = \left( \frac{x'}{a} \right)^2 + \left( \frac{y'}{b} \right)^2 \quad .
\end{equation}
In analogy with the EFF87 $\alpha$ parameter, which is related to the core radius, $a$ and $b$ are related to the core semi-major and semi-minor axes of the elliptically symmetric GEFF profile.
The quantities $x'$ and $y'$ are given by:
\begin{equation}
\left( \begin{array}{c} x' \\ y' \end{array} \right) = \left( \begin{array}{cc} \cos \theta & - \sin \theta \\ \sin \theta & \cos \theta \end{array}\right) \times \left( \begin{array}{c} x - x_{\rmn{c}} \\ y - y_{\rmn{c}} \end{array} \right)  \; ;
\end{equation}
where $(x_{\rmn{c}}, y_{\rmn{c}})$ are the pixel-coordinates of the centre of the ellipse and $\theta$ is the tilt angle between the semimajor axis and the x axis, measured counterclockwise.

Summarizing, a GEFF profile has 8 different parameters: $\mathbf{P} = (\Sigma_{\rmn{BG}}, \Sigma_{\rmn{c}}, \Gamma, a, b, x_{\rmn{c}}, y_{\rmn{c}}, \theta )$. In the fit we left all of them free, apart from the exponent $\Gamma$. Since the equivalent exponent for an EFF87 profile was found to be $\beta = 2$ in Paper I, we constrained our $\Gamma$ to stay between 1 and 3.

% In Paper I, the equivalent exponent for an EFF87 profile was found to be 2, corresponding to an ''isothermal'' system. Hence we constrained our $\Gamma$ to stay between 1 and 3.
Moreover, given that the completeness correction in the very centre of Wd~1 may be uncertain, we performed the fit by neglecting the region in which the stars have, on average, a completeness factor smaller than 0.25.
An example of stellar density contours and the relative GEFF fit is given in Fig. \ref{fig:dencon}.

\subsection{Results of the GEFF fit}

\begin{figure*}
 \centering
	\resizebox{\hsize}{!}{\includegraphics{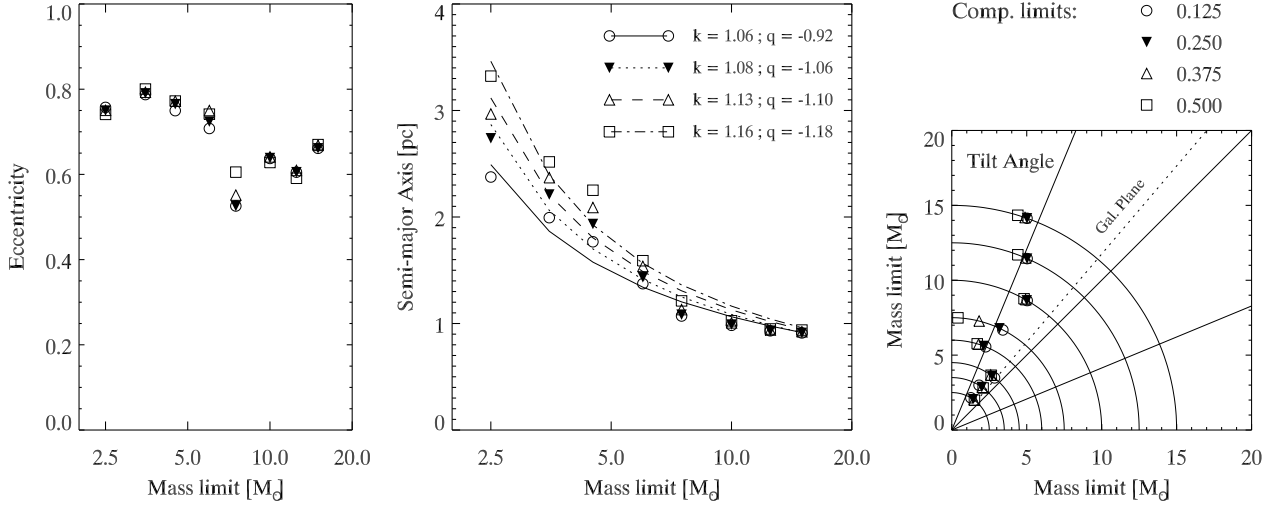}}
	\caption{\emph{Left}: Eccentricity values as a function of the minimum mass; \emph{Middle}: Semi-major axis values as a function of the minimum mass; the quoted $k$ and $q$ values are obtained by fitting functions as described in equation (\ref{eq:pwlaw}); \emph{Right}: Tilt angle between semi-major axis and x axis, measured counterclockwise; each of the concentric rings helps to distinguish the different values as a function of the minimum mass used. Different symbols correspond to different completeness thresholds.}
 \label{fig:shapepar}
\end{figure*}

We calculated the stellar density for several combinations of the mass and completeness thresholds; the values used are: $m [M_{\sun}] = [2.5,3.5,4.5,6.,7.5,10.,12.5,15.]$ and $\mathcal{C} = [0.125,0.25,0.375,0.5]$.
A summary of the outcome of the GEFF fit for all these combinations can be found in Fig. \ref{fig:shapepar}. For clarity, we emphasize that, given the value of the mass threshold, we consider all the stars with $m > m_{\rmn{low}}$. Hence, in the plots of Fig. \ref{fig:shapepar}, going from right to left along the mass axis, it is possible to see the \emph{cumulative} effect of including lower and lower mass stars.
The different symbols represent different completeness thresholds, as indicated.

In the left panel of the figure we show  the eccentricity, $\epsilon = \sqrt{ 1- \left(  \frac{b}{a} \right)^2}$. With this definition, the ellipticity values, $\eta = 0.15 \,\rmn{and} \,0.19$, of Paper I become $\epsilon = 0.53 \,\rmn{and}\, 0.59$, respectively.
The eccentricity values are almost constant with mass.  Their average values are somewhat higher than what found in Paper I, with $\epsilon \approx 0.75$ indicating an axis ratio $a:b = 3:2$. The fact that the numbers are slightly different compared to Paper I is not surprising. The adoption of a radially symmetric completeness correction  in Paper I has partially smoothed out some of the asymmetry and intrinsic elongation of the cluster. %The present work clearly strengthen the previous, very interesting finding that Wd~1 is indeed an elongated cluster.
%The eccentricity is slightly decreasing towards higher masses. It is hard to state whether this effect is real or is just a result of the uncertainties in the GEFF profile fit. Assuming that the effect is real, then the distribution of the most massive stars is slightly more isotropic than that of the less massive ones. 
Our new results clearly reveal the elongated 2D density distribution of main-sequence stars with masses between $\approx 3 \, \rmn{and} \, \approx 30 \,M_{\sun}$, with higher elongation observed for lower mass stars.
This might be related to the fact that more massive stars are also more centrally concentrated (see below), hence their average collision time is shorter than that of the less massive stars. Consequently, massive stars undergo more dynamical interactions and their momenta become more isotropic.

%As anticipated, the length of the semi-major axis is a decreasing function of the stellar mass. 
For completeness values between 0.125 and 0.5 we also performed a least-square fit of the semi-major axis values, using the functional relation:
\begin{equation}
\label{eq:pwlaw}
a = k \times \left[ \log(m) \right]^q \; .
\end{equation}
The results of the fits are shown in the central panel of Fig. \ref{fig:shapepar}. In general the cumulative semimajor axis decreases almost like $1/\log(m)$.
 We already found an indication of mass segregation by investigating the IMF spatial variations (see Sect. \ref{sec:IMF}). The finding here confirms that massive stars are more centrally concentrated. The figure also shows that adopting a lower completeness threshold, the actual size of the semi-major axis decreases, at fixed mass. The reason is the inclusion of more and more stars in the centre of the cluster, where, of course, the total completeness is lower. Consequently, going to lower completeness thresholds the stellar density has a more pronounced peak in the centre, while the density in the outer regions of the cluster does not change as much. Since $a$ is a measure of the length-scale of the density decay with distance from the centre, we obtain lower $a$ values when the density contrast between the centre and the outskirts is more pronounced.

From Fig. \ref{fig:shapepar}, right panel, it is also very interesting to note that Wd~1's direction of elongation lies very close to the galactic plane.
 %It is interesting to note that the semi-major axis of the ellipses are quite well aligned with the galactic plane (see Fig. \ref{fig:shapepar}, right panel). 
% The right panel of Fig. \ref{fig:shapepar} shows the angles between $a$ and the x axis; the dotted line is the direction of the galactic plane. In this polar plot the radial distances from the origin represent the values of the mass threshold.
%It is clear that the elongation direction of Wd~1 lies in the galactic plane or very close to it This provides a hint for potential causes of the Wd~1 elongation.
\label{subsec:resgeff}

\subsection{Possible sources of elongation}

%The discussion that follows will be qualitative, simulations are needed to address the problem of Wd~1 shape and density profile in detail. Nevertheless we illustrate some possible explanations, giving order of magnitude estimates.% and raising the challenge for experts of cluster dynamical evolution to find the best theoretical explanation to our observational findings.
In the following, we carry out a qualitative discussion of possible sources of elongation.
The typical orbital period for a star at 1pc distance from a central point source with mass $M=10^5 M_{\sun}$ is about $t_d \approx 3\times 10^5$ Myr. This timescale is much shorter than the typical half-mass relaxation time of about $t_{rh} \approx 10^8 - 10^9$ yr, as determined for a typical globular cluster --comparable in stellar mass to Wd~1--, and defined as the time required for the central half of the cluster mass to reach equilibrium \citep[see Sect. 1.1 in][]{Spitzer:1987lr}. Given the difference in the two timescales, it is clear that the observed deviation from spherical symmetry cannot be ascribed to the global evolution of the cluster as an \emph{isolated} system; after few orbits and encounters, the phase-space distribution of stars for an isolated system is expected to be isotropic (in $\mathbf{v}$) and spherically symmetric (in $\mathbf{r}$).
The deviation from the spherical cluster shape must be explained either by unusual initial conditions still reflected in the present cluster appearance or by some interaction with the rest of the Galaxy.\\
%Regarding the initial conditions, we tentatively propose two scenarios. 
%One would be the presence of a net angular momentum in the giant molecular cloud from which Wd~1 has formed. Given its location in the Scutum-Crux arm,  differential-rotation-induced shear could be responsible for this net momentum. Hence the Wd~1 shape could be explained as a rotation ellipsoid with rotation axis perpendicular to the galactic plane. This, for a situation where the angular momentum has been conserved in the contraction phase of the cloud and during the Wd~1 star formation phase. \cite{Ballesteros-Paredes:2009lr} study the balance of internal gravitational forces and galactic tidal and shearing effects on a molecular cloud as a function of its initial elongation. They find that the more elongated clouds suffer a stronger external influence with shear and tides acting against the self gravity of the clouds, eventually quenching star formation and disrupting them. These findings may rule out differential rotation as the primary source of elongation of Wd~1.\\
A net angular momentum of the giant molecular cloud forming Wd~1 or a formation of Wd~1 out of two or more subclusters might either be responsible for its elongated shape.
Differential galactic rotation exerts a shear on molecular clouds which might lead to a net angular momentum. According to a recent study by \cite{Ballesteros-Paredes:2009lr}, galactic shear and tides have rather strong effects on initially elongated clouds, eventually quenching star formation and disrupting the clouds. Hence differential rotation is an unlikely source for the elongation of Wd~1.\\
An other intriguing possibility would be a ''hierarchical'' formation scenario, with merging of two or more smaller subclusters. The existence of a non-negligible fraction of possible binary clusters is supported by observations \citep[see e.g.][]{de-la-Fuente-Marcos:2008av}. A hierarchical organization of the ISM and of young stellar groups and clusters is indeed observed on many scales \citep[][]{Elmegreen:2009lr}. 
\cite{Negueruela:2010fj} report  the presence of a subclump of massive stars in the S-E region of Wd~1, even though they also warn that fluctuations in the star counts could be responsible for this observed clump.
\cite{Clark:2005lr} suggest that an age spread within the Wd~1 massive star population is really unlikely, hence any possible merging or capture event must have happened in the very beginning of the cluster's history. Otherwise this episode could have happened also more recently, but under the condition that the subclusters are coeval, i.e. star formation has occurred at almost the same time in different regions of the giant molecular cloud.
If this scenario would be true, the modest amount of dynamical crossing times occurred from Wd~1 formation (age $\simeq 10\, t_d$) may be the reason why the stellar motion have not yet reached an isotropic distribution.
No dynamical simulations that include such a large number of particles as Wd~1 members have been performed so far. Nevertheless, recent studies suggest that merging is indeed possible over a wide range of initial conditions \citep[see][and references therein]{Portegies-Zwart:2007el,de-la-Fuente-Marcos:2010fk}.\\
%An energetic event that could provide enough total energy to perturbed the stellar motions could be a supernova explosion.
%As noted before, some $\sim 65$ supernovae are expected to have occurred in Wd~1. Anyway it remains questionable how this energy could be transferred by the ejecta to the stars. There is quite an understanding about the ejecta-intracluster medium interaction that depletes the cluster from its remaining gas reservoir from the initial star formation event, but it seems unlikely that supernovae ejecta could perturb stellar motion. An other possibility would be that the anisotropies of supernovae explosions
We evaluate whether tidal effects in the Galactic central field could be responsible for the Wd~1 shape. Under the simplifying assumption of a circular restricted 3-body problem, with primary mass at the Galactic Centre position and secondary at  the cluster centre, we find that $r_L$, i.e. the distance of the inner Lagrangian point from Wd~1 centre is:
$$
{r_L}^3 = \frac{1}{3}\frac{M_{\rmn{Wd\,1}}}{M_{\rmn{MW}}} {R_G}^3 \, ; 
$$
where $M_{\rmn{Wd\,1}} =  5\times 10^4 M_{\sun}$ is our Wd~1 mass estimate, $M_{\rmn{MW}} = 6 \times 10^{11} M_{\sun}$ is the mass of the Milky Way and $R_G = 4 \,\rmn{kpc}$ is the Wd~1 galactocentric distance.
With these numbers, we obtain $r_L = 12 \,\rmn{pc}$; given that $M_{\rmn{Wd\,1}} \ll M_{\rmn{MW}} $ this is also the distance for the external Lagrangian point from the Wd~1 centre. This estimate for the tidal radius is a lower limit, since it was assumed that all the mass of the Galaxy is confined within the orbit of Wd1. A more correct estimate, taking into account only the enclosed galactic mass at radius $R_G$, would lead to an even larger value of $r_L$. Consequently, it is clear that Wd~1 is far from filling its Roche lobe, while we measure elongation already on a scale of $\sim 1 \, \rmn{pc}$ from the cluster core. Hence tidal distortion from the whole Galaxy is unlikely the reason for the elongation of Wd~1.\\
An other important tidal effect could be caused by the galactic disk gravitational field. The disk potential is costant far away from the galactic midplane, where the matter density distribution of the disk drops to zero. On the contrary within the disk the potential has a non-zero divergence in the direction perpendicular to the plane (Z axis). This divergence causes a net acceleration of the stellar motions in the Z direction, and, as a result, an initially spherical and isotropic cluster moving across the disk midplane is \emph{compressed} along the Z direction. This phenomenon is known as ''compressive gravitational shock'' and an analytical solution is presented in Chapter 5 by \cite{Spitzer:1987lr}. Unfortunately the conditions that are required to apply this analytical treatment do not hold for Wd~1. In \cite{Spitzer:1987lr} the galactic plane tidal field is treated as a fast time-dependent perturbation to the motion of stars within Globular Clusters which cross the midplane at a speed of hundreds of km/s. Hence the duration of the perturbation is short compared to the stellar orbital period around the cluster centre. However, Wd~1 is moving much slower in the Z direction. From a preliminary analysis of our AO multi-epoch observation, we can set a limit on the net bulk motion of $\la 10$ km/s along the galactic longitude coordinate $b$ (Kudryavtseva et al. \emph{in preparation}).
For this reason the \cite{Spitzer:1987lr} analytical solution cannot be used here, but it could be worth to investigate the effects of the disk tidal field in detail with dynamical simulations.
%Nevertheless, we believe that the tidal force from the galactic disk could be responsible for the elongation of Wd~1, hence it would be worth to investigate this scenario in detail with dynamical simulations.

\subsection{The effective cumulative radius}

As an alternative to study of mass segregation, in addition to the estimate of the GEFF best fit semi-major axis, $a$, we used an independent quantity that we call the effective cumulative radius:
\begin{equation}
 r_{\rmn{eff}} (m) = \sqrt \frac{ \sum_{m_i > m} \left( \frac{r_i}{\mathcal{C}_{J\,i} \times \mathcal{C}_{K_{\rmn{S}}\,i}} \right)^2}{\sum_{m_i > m} \left( \frac{1}{\mathcal{C}_{J\,i}\times \mathcal{C}_{K_{\rmn{S}}\,i}}\right)^2} \quad .
\end{equation}
This quantity is obtained by taking all the stars with mass $m_i > m$ and computing their geometric-averaged distance from the centre of the cluster.
The distance from the centre for the single stars is $r_i$; the completeness factors, $\mathcal{C}_i$ are needed to take into account, in a statistical sense, the undetected sources.

\begin{figure}
 \centering
	\resizebox{\hsize}{!}{\includegraphics{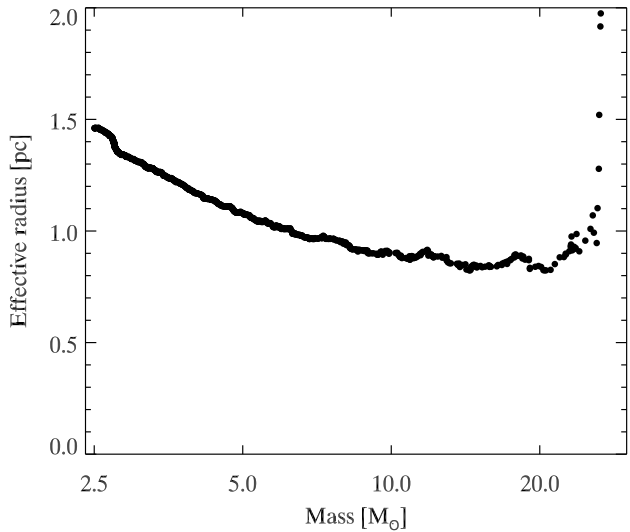}}
	\caption{Effective cumulative radius for Wd~1 stars. Note that the $\sim 10$ rightmost points are affected by the off-centre location of the one most massive star in our data.}
 \label{fig:effrad}
\end{figure}

Figure \ref{fig:effrad} shows the quantity $r_{\rmn{eff}}$ as a function of mass. Looking from right to left it is evident that the inclusion of less and less massive stars in the computation of $r_{\rmn{eff}}$ leads to an increase of the average distance from the centre of the cluster, meaning that more massive stars are located on average at  smaller effective radii compared to less massive stars. The first few points on the right of the plot don't follow this relation. This is due to the fact that the definition of an \emph{average} distance for the most massive stars is problematic, given the low numbers considered. Indeed, the most massive star in our data set is a bit off-centre, hence the effective radius for this star is quite large; this star also affects the effective radius of the first $\sim 10$ points on the right of the diagram, because its distance from the centre enters the computation of the average distance for the other stars. As long as the number of stars included in the computation of $r_{\rmn{eff}}$ increases, the results converge towards a more stable averaged distance.

The smooth increase of $r_{\rmn{eff}}$ with decreasing mass confirms the findings for the semi-major axis length of Sect. \ref{subsec:resgeff}. Hence we can state that Wd~1 is clearly mass segregated. 

\subsection{The origin of mass segregation for Wd 1}

Mass segregation is a phenomenon observed in many young clusters; some examples are the Orion Nebula Cluster \citep{Hillenbrand:1998lr}, the NGC 3603 Young Cluster \citep{Stolte:2006kx,Harayama:2008yq}, the Arches Cluster \citep{Stolte:2005vn,Kim:2006rt,Espinoza:2009ys}, and debate is still open whether this phenomenon is either primordial or due to dynamical evolution.
If only two-body encounters are considered, the half-mass relaxation time for a cluster is given by \citep{Binney:1987qy}:
%t_{\rmn{rel,2b}} \approx \frac{N}{8 \ln N} t_c \, .
$$
t_{\rmn{rh}} = \frac{6.5 \times 10^8 \, \rmn{yr}}{\ln(0.4 \,N)} \left(\frac{M}{10^5 M_{\sun}}\right)^{\frac{1}{2}} \left(\frac{ 1 M_{\sun}}{<m>}\right) \left(\frac{r_{\rmn{h}}}{1 \,\rmn{pc}}\right)^{\frac{3}{2}}
$$
where $N$ is the total number of stars, $<m>$ is the mean stellar mass and $r_{\rmn{h}}$ the deprojected half-mass radius, equal to 4/3 of the projected half-mass radius.
Considering $N = 10^5$, $M = 5\times10^4 M_{\sun}$, $r_{\rmn{h}} = 4/3 \times 1.1$ pc (see Paper I) and $<m> = 0.6 M_{\sun}$, we get an estimate of $t_{\rmn{rh}} \approx 130$ Myr, much larger than the age of the cluster. This discrepancy between relaxation time and age is common to many clusters and has been used as an argument  in favor of the primordial segregation scenario \citep{Bonnell:1998fk}.
Nevertheless one has to consider that the time for a star with mass $m_*$ to drift towards the cluster centre due to dynamical friction is:
$$
t_{\rmn{df}} = \frac{<m>}{m_*} \times t_{\rmn{rh}}
$$
In the case of Wd~1 this segregation time can be as short as 2\% of $t_{\rmn{rh}}$, i.e. 2.6 Myr for a star of $\sim30 M_{\sun}$, i.e. the most massive stars in our sample.
Hence primordial segregation would not be necessary to explain the observed mass segregation. Furthermore the evolution of Wd~1  has probably been more complex, and the value of $t_{\rmn{rh}}$ might have changed in time.
Mass loss from stellar winds, supernovae explosions and gas removal might have caused a global expansion of Wd~1, meaning that $t_{\rmn{rh}}$ was shorter in the past \citet{Gieles2:2010fk}. \cite{Gurkan:2004fk} and \cite{Portegies-Zwart:2004lr} showed that the core collapse time for massive clusters is about 0.1-0.2 $t_{\rmn{rh}}$. It is therefore not unlikely that the core of Wd~1 has undergone a dynamical collapse, which is then also followed by  expansion that could increase the relaxation time. This would push down the mass limit which we expect to be affected by mass segregation.
\cite*{McMillan:2007fk} have proposed an alternative scenario to the primordial segregation one, in order to explain mass segregation observed in young cluster. This scenario predicts that mass-segregated young, massive clusters could be the product of merging of several smaller subclusters. Substructure in molecular clouds is observed in both density and kinematics \citep[][]{Williams:1999lr,Williams:2000fk} and this substructure is reflected as well in young clusters \citep[][]{Larson:1995qy,Testi:2000uq,Gutermuth:2005kx,Allen:2007yq}.
In a hierarchical formation scenario, a massive cluster could be formed by the merging of several subclusters and still might show mass segregation.
Given their smaller $N$, the subclusters can rapidly reach a mass segregated status before they merge and regardless of the initial spatial distribution of stars. The mass segregation in these smaller clusters is favored by shorter $t_{\rmn{rh}}$. In addition, \cite{Allison:2009lr,Allison:2010fk} show that  clusters may undergo an initial collapse phase which can significantly accelerate mass segregation. In this phase the cluster forms a very dense and clumpy core, where the massive stars can rapidly segregate given the short crossing time and large number of encounters. The timescale for such a process in a cluster with $N \sim 10^3$ is $\,\la 1 \rmn{Myr}$. The simulations by \cite*{McMillan:2007fk} additionally show that mass segregation developed by single subclusters is preserved during merging. Consequently the final massive cluster exhibits mass segregation at an age much smaller than its global relaxation time.  While merging  might not be required to explain mass segregation, it could also explain the elongated shape of Wd~1. Hence we think that this is a very interesting scenario for Wd~1 formation. % because {\bf in addition to speeding up the mass segregation process} it could also be an explanation for its elongated shape (see Sect. \ref{subsec:resgeff}). 
We point out that \cite{McMillan:2007fk} have carried out pure N-body simulations. Recently, \cite{Bate:2009lr} has performed hydrodynamic simulation of star-forming regions that include gas drag and gas accretion onto stars, in addition to the mutual gravity between them. The final cluster that is formed is the result of the merging of 5 subclusters. The author finds no evidence for mass segregation. However the number of stars formed in his simulations is of the order of $10^3$ with the most massive star of only $\approx 5 M_{\sun}$. Hence the simulated cluster cannot be directly compared to Wd~1.
More recently, using the final stellar positions of \cite{Bate:2009lr}, and applying their own segregation detection method, \cite{Moeckel:2009qy} found evidence of segregation at least for the most massive stars. This last scenario, in which only the few most massive members are found in the cluster's core, is more similar to what is observed for the Trapezium stars in the ONC, than to what we observe in Wd~1, i.e. an evidence of continuous mass segregation across the whole stellar mass spectrum.

%On the extreme left of the plot $r_{\rmn{eff}}$ reaches a plateau. This is not intrinsic, but simply due to the fact that there are no stars measured \emph{outside} the frame. To study the spatial distribution of even lower mass stars in a more precise way a wider FoV image of Wd~1 would be needed.

\section{Conclusions}
We have presented a new, thorough analysis of near-infrared data for the intermediate- and low-mass stellar population of the massive young cluster Westerlund 1. Using artificial stars, we have been able to sample spatial variations of photometric completeness on a scale of few stellar FWHM . The same artificial stars have been used to infer realistic photometric error estimates, as well as the correlation between errors in different bands.
Incompleteness corrections and errors were used to apply a novel statistical field subtraction technique to the data. Using a nearby control field we obtained a clean sample of cluster members.
The clean catalogue of stars, together with state-of-the-art stellar models, has been used to determine the cluster's properties. We derived an extinction $A_{K_{\rmn{S}}} = 0.91 \pm 0.05 \,\rmn{mag}$, an age $\tau = 4 \pm 0.5$ Myr and a distance $d = 4.0 \pm 0.2 \,\rmn{kpc}$.

We investigated the cluster's IMF slope using a new approach to stellar mass determination. The information on magnitude errors and their correlation has been used to derive the mass-probability-distribution for each star, given the best-fit isochrone.
The completeness-corrected IMF has a slope of $\gamma = 2.44_{-0.08}^{+0.20}$, slightly steeper than the Salpeter or Kroupa IMF; this slight discrepancy could be partially reconciled if we consider that, for the sake of simplicity, we are neglecting the influence of (unknown) undetected binarity, hence our quoted error is probably an underestimate of the total, statistical plus systematic error \citep{2009Ap&SS.324...95M}.
From the IMF slope and its normalization constant we found a total mass for the cluster of $M_{\rmn{Wd \, 1}} = 4.91_{-0.49}^{+1.79}  \times 10^4 M_{\sun}$.

The spatially varying completeness, combined with the probabilistic mass determination, enabled us to investigate the spatial variations of the IMF. The Wd~1 starburst cluster is mass segregated, with massive stars more centrally concentrated.
Other indications of mass segregation come from the analysis of the stellar density distribution.
In order to study the 2D density distribution as a function of stellar mass, we fitted 2D elliptical profiles.
This analisys revealed a tight dependency of the ellipses semi-major axis length on mass:$a(m) \propto 1/\log(m)$. 
Given the young age of Wd~1, its global mass segregation cannot be explained by simple 2-body relaxation.
Interestingly, from the density distribution analysis, we found that Wd~1 is elongated along the Galactic Plane with an axis ratio $a:b = 3:2$.
The mass segregation and the elongation together hint at a formation scenario involving the merging of multiple subclusters formed almost coevally in the parental giant molecular cloud.

\section*{Acknowledgments}

The authors would like to thank E. Tognelli (Universita' di Pisa, Pisa, Italy) for his support in the calculation of the Pisa stellar models.
We made large use of the IDL programming language, adopting the \emph{The IDL Astronomy User's Library}, the \emph{Coyote Library} and the \emph{MPFIT} package. MG would like to thank  the developers and maintainers of this very useful tools for making them publically available.
MG would also like to thank M. Fang, B. Vaidya and R. Andrae (all at Max-Planck-Institut f\"{u}r Astronomie, Heidelberg, Germany) for useful discussions.

%\bibliography{Wd1_gennaro}{}

\begin{thebibliography}{}

\bibitem[\protect\citeauthoryear{{Allen}, {Megeath}, {Gutermuth}, {Myers},
  {Wolk}, {Adams}, {Muzerolle}, {Young} \& {Pipher}}{{Allen}
  et~al.}{2007}]{Allen:2007yq}
{Allen} L.,  {Megeath} S.~T.,  {Gutermuth} R.,  {Myers} P.~C.,  {Wolk} S.,
  {Adams} F.~C.,  {Muzerolle} J.,  {Young} E.,    {Pipher} J.~L.,  2007, B.
  Reipurth , D. Jewitt, K. Keil, eds., Protostars and Planets V University, of
  Arizona Press, Tucson, p.~361

\bibitem[\protect\citeauthoryear{{Allison}, {Goodwin}, {Parker}, {de Grijs},
  {Portegies Zwart} \& {Kouwenhoven}}{{Allison} et~al.}{2009}]{Allison:2009lr}
{Allison} R.~J.,  {Goodwin} S.~P.,  {Parker} R.~J.,  {de Grijs} R.,  {Portegies
  Zwart} S.~F.,    {Kouwenhoven} M.~B.~N.,  2009, \apjl, 700, L99

\bibitem[\protect\citeauthoryear{{Allison}, {Goodwin}, {Parker}, {Portegies
  Zwart} \& {de Grijs}}{{Allison} et~al.}{2010}]{Allison:2010fk}
{Allison} R.~J.,  {Goodwin} S.~P.,  {Parker} R.~J.,  {Portegies Zwart} S.~F.,
   {de Grijs} R.,  2010, \mnras, 407, 1098

\bibitem[\protect\citeauthoryear{{Andersen}, {Zinnecker}, {Moneti},
  {McCaughrean}, {Brandl}, {Brandner}, {Meylan} \& {Hunter}}{{Andersen}
  et~al.}{2009}]{Andersen:2009xt}
{Andersen} M.,  {Zinnecker} H.,  {Moneti} A.,  {McCaughrean} M.~J.,  {Brandl}
  B.,  {Brandner} W.,  {Meylan} G.,    {Hunter} D.,  2009, \apj, 707, 1347

\bibitem[\protect\citeauthoryear{{Andrae}}{{Andrae}}{2010}]{Andrae:2010lr}
{Andrae} R.,  2010, ArXiv:1009.2755v3

\bibitem[\protect\citeauthoryear{{Ballesteros-Paredes}, {G{\'o}mez}, {Pichardo}
  \& {V{\'a}zquez-Semadeni}}{{Ballesteros-Paredes}
  et~al.}{2009}]{Ballesteros-Paredes:2009lr}
{Ballesteros-Paredes} J.,  {G{\'o}mez} G.~C.,  {Pichardo} B.,
  {V{\'a}zquez-Semadeni} E.,  2009, \mnras, 393, 1563

\bibitem[\protect\citeauthoryear{{Bate}}{{Bate}}{2009}]{Bate:2009lr}
{Bate} M.~R.,  2009, \mnras, 392, 590

\bibitem[\protect\citeauthoryear{{Binney} \& {Tremaine}}{{Binney} \&
  {Tremaine}}{1987}]{Binney:1987qy}
{Binney} J.,  {Tremaine} S.,  1987, {Galactic dynamics, Princeton University
  Press, Princeton}

\bibitem[\protect\citeauthoryear{{Bonnell} \& {Davies}}{{Bonnell} \&
  {Davies}}{1998}]{Bonnell:1998fk}
{Bonnell} I.~A.,  {Davies} M.~B.,  1998, \mnras, 295, 691

\bibitem[\protect\citeauthoryear{{Brandner}, {Clark}, {Stolte}, {Waters},
  {Negueruela} \& {Goodwin}}{{Brandner} et~al.}{2008}]{Brandner:2008lr}
{Brandner} W.,  {Clark} J.~S.,  {Stolte} A.,  {Waters} R.,  {Negueruela} I.,
  {Goodwin} S.~P.,  2008, \aap, 478, 137

\bibitem[\protect\citeauthoryear{{Brott} \& {Hauschildt}}{{Brott} \&
  {Hauschildt}}{2005}]{2005ESASP.576..565B}
{Brott} I.,  {Hauschildt} P.~H.,  2005, in {Turon} C.,  {O'Flaherty} K.~S.,
  {Perryman} M.~A.~C.,  eds, The Three-Dimensional Universe with Gaia, Vol.~576
  of ESA Special Publication, {A PHOENIX Model Atmosphere Grid for Gaia}.
p.~565

\bibitem[\protect\citeauthoryear{{Campbell}, {Evans}, {Mackey}, {Gieles},
  {Alves}, {Ascenso}, {Bastian} \& {Longmore}}{{Campbell}
  et~al.}{2010}]{Campbell:2010hc}
{Campbell} M.~A.,  {Evans} C.~J.,  {Mackey} A.~D.,  {Gieles} M.,  {Alves} J.,
  {Ascenso} J.,  {Bastian} N.,    {Longmore} A.~J.,  2010, \mnras, 405, 421

\bibitem[\protect\citeauthoryear{{Castelli} \& {Kurucz}}{{Castelli} \&
  {Kurucz}}{2003}]{2003IAUS..210P.A20C}
{Castelli} F.,  {Kurucz} R.~L.,  2003, in {Piskunov} N.,  {Weiss} W.~W.,
  {Gray} D.~F.,  eds, Modelling of Stellar Atmospheres, Vol.~210 of IAU
  Symposium, {New Grids of ATLAS9 Model Atmospheres}.
p.~20

\bibitem[\protect\citeauthoryear{{Clark}, {Negueruela}, {Crowther} \&
  {Goodwin}}{{Clark} et~al.}{2005}]{Clark:2005lr}
{Clark} J.~S.,  {Negueruela} I.,  {Crowther} P.~A.,    {Goodwin} S.~P.,  2005,
  \aap, 434, 949

\bibitem[\protect\citeauthoryear{{Crowther}, {Hadfield}, {Clark}, {Negueruela}
  \& {Vacca}}{{Crowther} et~al.}{2006}]{Crowther:2006fk}
{Crowther} P.~A.,  {Hadfield} L.~J.,  {Clark} J.~S.,  {Negueruela} I.,
  {Vacca} W.~D.,  2006, \mnras, 372, 1407

\bibitem[\protect\citeauthoryear{{de la Fuente Marcos} \& {de la Fuente
  Marcos}}{{de la Fuente Marcos} \& {de la Fuente
  Marcos}}{2008}]{de-la-Fuente-Marcos:2008av}
{de la Fuente Marcos} R.,  {de la Fuente Marcos} C.,  2008, \apj, 672, 342

\bibitem[\protect\citeauthoryear{{de la Fuente Marcos} \& {de la Fuente
  Marcos}}{{de la Fuente Marcos} \& {de la Fuente
  Marcos}}{2010}]{de-la-Fuente-Marcos:2010fk}
{de la Fuente Marcos} R.,  {de la Fuente Marcos} C.,  2010, \apj, 719, 104

\bibitem[\protect\citeauthoryear{{Degl'Innocenti}, {Prada Moroni}, {Marconi} \&
  {Ruoppo}}{{Degl'Innocenti} et~al.}{2008}]{DeglInnocenti:2008fk}
{Degl'Innocenti} S.,  {Prada Moroni} P.~G.,  {Marconi} M.,    {Ruoppo} A.,
  2008, \apss, 316, 25

\bibitem[\protect\citeauthoryear{{Devillard}}{{Devillard}}{2001}]{Devillard:20%
01fk}
{Devillard} N.,  2001, in {F.~R.~Harnden Jr., F.~A.~Primini, \& H.~E.~Payne}
  ed., Astronomical Data Analysis Software and Systems X, Vol.~238 of
  Astronomical Society of the Pacific Conference Series, {ESO C Library for an
  Image Processing Software Environment (eclipse)}.
p.~525

\bibitem[\protect\citeauthoryear{{Efron}}{{Efron}}{1979}]{Efron:1979lk}
{Efron} B.,  1979, Ann. Statist., 7, 1

\bibitem[\protect\citeauthoryear{{Eisenhauer}, {Quirrenbach}, {Zinnecker} \&
  {Genzel}}{{Eisenhauer} et~al.}{1998}]{Eisenhauer:1998lr}
{Eisenhauer} F.,  {Quirrenbach} A.,  {Zinnecker} H.,    {Genzel} R.,  1998,
  \apj, 498, 278

\bibitem[\protect\citeauthoryear{{Elmegreen}}{{Elmegreen}}{2009}]{Elmegreen:20%
09lr}
{Elmegreen} B.~G.,  2009, in {Richtler, T.~\& Larsen, S.} ed., Globular
  Clusters - Guides to Galaxies, {Hierarchical Formation of Galactic Clusters}.
p.~87

\bibitem[\protect\citeauthoryear{{Elson}, {Fall} \& {Freeman}}{{Elson}
  et~al.}{1987}]{Elson:1987lr}
{Elson} R.~A.~W.,  {Fall} S.~M.,    {Freeman} K.~C.,  1987, \apj, 323, 54

\bibitem[\protect\citeauthoryear{{Espinoza}, {Selman} \& {Melnick}}{{Espinoza}
  et~al.}{2009}]{Espinoza:2009ys}
{Espinoza} P.,  {Selman} F.~J.,    {Melnick} J.,  2009, \aap, 501, 563

\bibitem[\protect\citeauthoryear{{Fleck}, {Boily}, {Lan{\c c}on} \&
  {Deiters}}{{Fleck} et~al.}{2006}]{Fleck:2006qy}
{Fleck} J.,  {Boily} C.~M.,  {Lan{\c c}on} A.,    {Deiters} S.,  2006, \mnras,
  369, 1392

\bibitem[\protect\citeauthoryear{{Gieles}, {Sana} \& {Portegies
  Zwart}}{{Gieles} et~al.}{2010a}]{Gieles:2010fj}
{Gieles} M.,  {Sana} H.,    {Portegies Zwart} S.~F.,  2010a, \mnras, 402, 1750

\bibitem[\protect\citeauthoryear{{Gieles}, {Baumgardt}, {Heggie} \&
  {Lamers}}{{Gieles} et~al.}{2010b}]{Gieles2:2010fk}
{Gieles} M.,  {Baumgardt} H.,  {Heggie} D.~C.,    {Lamers} H.~J.~G.~L.~M.,
  2010b, \mnras, 408, L16

\bibitem[\protect\citeauthoryear{{G{\"u}rkan}, {Freitag} \&
  {Rasio}}{{G{\"u}rkan} et~al.}{2004}]{Gurkan:2004fk}
{G{\"u}rkan} M.~A.,  {Freitag} M.,    {Rasio} F.~A.,  2004, \apj, 604, 632

\bibitem[\protect\citeauthoryear{{Gutermuth}, {Megeath}, {Pipher}, {Williams},
  {Allen}, {Myers} \& {Raines}}{{Gutermuth} et~al.}{2005}]{Gutermuth:2005kx}
{Gutermuth} R.~A.,  {Megeath} S.~T.,  {Pipher} J.~L.,  {Williams} J.~P.,
  {Allen} L.~E.,  {Myers} P.~C.,    {Raines} S.~N.,  2005, \apj, 632, 397

\bibitem[\protect\citeauthoryear{{Harayama}, {Eisenhauer} \&
  {Martins}}{{Harayama} et~al.}{2008}]{Harayama:2008yq}
{Harayama} Y.,  {Eisenhauer} F.,    {Martins} F.,  2008, \apj, 675, 1319

\bibitem[\protect\citeauthoryear{{Hastie}, {Tibshirani} \& {Friedman}}{{Hastie}
  et~al.}{2009}]{Hastie2009ji}
{Hastie} T.,  {Tibshirani} R.,    {Friedman} J.,  2009, The Elements of
  Statistical Learning: Data Mining, Inference, and Prediction, Springer
  Science+Business Media, New York

\bibitem[\protect\citeauthoryear{{Hillenbrand} \& {Hartmann}}{{Hillenbrand} \&
  {Hartmann}}{1998}]{Hillenbrand:1998lr}
{Hillenbrand} L.~A.,  {Hartmann} L.~W.,  1998, \apj, 492, 540

\bibitem[\protect\citeauthoryear{{Kim}, {Figer}, {Kudritzki} \&
  {Najarro}}{{Kim} et~al.}{2006}]{Kim:2006rt}
{Kim} S.~S.,  {Figer} D.~F.,  {Kudritzki} R.~P.,    {Najarro} F.,  2006, \apjl,
  653, L113

\bibitem[\protect\citeauthoryear{{Kothes} \& {Dougherty}}{{Kothes} \&
  {Dougherty}}{2007}]{Kothes:2007fk}
{Kothes} R.,  {Dougherty} S.~M.,  2007, \aap, 468, 993

\bibitem[\protect\citeauthoryear{{Kouwenhoven} \& {de Grijs}}{{Kouwenhoven} \&
  {de Grijs}}{2008}]{Kouwenhoven:2008uq}
{Kouwenhoven} M.~B.~N.,  {de Grijs} R.,  2008, \aap, 480, 103

\bibitem[\protect\citeauthoryear{{Kroupa}}{{Kroupa}}{2001}]{2001MNRAS.322..231%
K}
{Kroupa} P.,  2001, \mnras, 322, 231

\bibitem[\protect\citeauthoryear{{Larson}}{{Larson}}{1995}]{Larson:1995qy}
{Larson} R.~B.,  1995, \mnras, 272, 213

\bibitem[\protect\citeauthoryear{{Lejeune} \& {Schaerer}}{{Lejeune} \&
  {Schaerer}}{2001}]{2001A&A...366..538L}
{Lejeune} T.,  {Schaerer} D.,  2001, \aap, 366, 538

\bibitem[\protect\citeauthoryear{{Ma{\'{\i}}z Apell{\'a}niz}}{{Ma{\'{\i}}z
  Apell{\'a}niz}}{2009}]{2009Ap&SS.324...95M}
{Ma{\'{\i}}z Apell{\'a}niz} J.,  2009, \apss, 324, 95

\bibitem[\protect\citeauthoryear{{Marigo}, {Girardi}, {Bressan}, {Groenewegen},
  {Silva} \& {Granato}}{{Marigo} et~al.}{2008}]{Marigo:2008lr}
{Marigo} P.,  {Girardi} L.,  {Bressan} A.,  {Groenewegen} M.~A.~T.,  {Silva}
  L.,    {Granato} G.~L.,  2008, \aap, 482, 883

\bibitem[\protect\citeauthoryear{{McMillan}, {Vesperini} \& {Portegies
  Zwart}}{{McMillan} et~al.}{2007}]{McMillan:2007fk}
{McMillan} S.~L.~W.,  {Vesperini} E.,    {Portegies Zwart} S.~F.,  2007, \apjl,
  655, L45

\bibitem[\protect\citeauthoryear{{Mengel} \& {Tacconi-Garman}}{{Mengel} \&
  {Tacconi-Garman}}{2009}]{Mengel:2009uq}
{Mengel} S.,  {Tacconi-Garman} L.~E.,  2009, \apss, 324, 321

\bibitem[\protect\citeauthoryear{{Meynet} \& {Maeder}}{{Meynet} \&
  {Maeder}}{2000}]{Meynet:2000qy}
{Meynet} G.,  {Maeder} A.,  2000, \aap, 361, 101

\bibitem[\protect\citeauthoryear{{Moeckel} \& {Bonnell}}{{Moeckel} \&
  {Bonnell}}{2009}]{Moeckel:2009qy}
{Moeckel} N.,  {Bonnell} I.~A.,  2009, \mnras, 400, 657

\bibitem[\protect\citeauthoryear{{Muno}, {Law}, {Clark}, {Dougherty}, {de
  Grijs}, {Portegies Zwart} \& {Yusef-Zadeh}}{{Muno}
  et~al.}{2006}]{2006ApJ...650..203M}
{Muno} M.~P.,  {Law} C.,  {Clark} J.~S.,  {Dougherty} S.~M.,  {de Grijs} R.,
  {Portegies Zwart} S.,    {Yusef-Zadeh} F.,  2006, \apj, 650, 203

\bibitem[\protect\citeauthoryear{{Negueruela} \& {Clark}}{{Negueruela} \&
  {Clark}}{2005}]{Negueruela:2005lr}
{Negueruela} I.,  {Clark} J.~S.,  2005, \aap, 436, 541

\bibitem[\protect\citeauthoryear{{Negueruela}, {Clark} \&
  {Ritchie}}{{Negueruela} et~al.}{2010}]{Negueruela:2010fj}
{Negueruela} I.,  {Clark} J.~S.,    {Ritchie} B.~W.,  2010, \aap, 516, A78

\bibitem[\protect\citeauthoryear{{Nishiyama}, {Nagata}, {Kusakabe},
  {Matsunaga}, {Naoi}, {Kato}, {Nagashima}, {Sugitani}, {Tamura}, {Tanab{\'e}}
  \& {Sato}}{{Nishiyama} et~al.}{2006}]{2006ApJ...638..839N}
{Nishiyama} S.,  {Nagata} T.,  {Kusakabe} N.,  {Matsunaga} N.,  {Naoi} T.,
  {Kato} D.,  {Nagashima} C.,  {Sugitani} K.,  {Tamura} M.,  {Tanab{\'e}} T.,
   {Sato} S.,  2006, \apj, 638, 839

\bibitem[\protect\citeauthoryear{{Nishiyama}, {Tamura}, {Hatano}, {Kato},
  {Tanab{\'e}}, {Sugitani} \& {Nagata}}{{Nishiyama}
  et~al.}{2009}]{Nishiyama:2009lr}
{Nishiyama} S.,  {Tamura} M.,  {Hatano} H.,  {Kato} D.,  {Tanab{\'e}} T.,
  {Sugitani} K.,    {Nagata} T.,  2009, \apj, 696, 1407

\bibitem[\protect\citeauthoryear{{Piatti}, {Bica} \& {Claria}}{{Piatti}
  et~al.}{1998}]{Piatti:1998qy}
{Piatti} A.~E.,  {Bica} E.,    {Claria} J.~J.,  1998, \aaps, 127, 423

\bibitem[\protect\citeauthoryear{{Plummer}}{{Plummer}}{1911}]{Plummer:1911mr}
{Plummer} H.~C.,  1911, \mnras, 71, 460

\bibitem[\protect\citeauthoryear{{Portegies Zwart}, {Baumgardt}, {Hut},
  {Makino} \& {McMillan}}{{Portegies Zwart}
  et~al.}{2004}]{Portegies-Zwart:2004lr}
{Portegies Zwart} S.~F.,  {Baumgardt} H.,  {Hut} P.,  {Makino} J.,
  {McMillan} S.~L.~W.,  2004, \nat, 428, 724

\bibitem[\protect\citeauthoryear{{Portegies Zwart} \& {Rusli}}{{Portegies
  Zwart} \& {Rusli}}{2007}]{Portegies-Zwart:2007el}
{Portegies Zwart} S.~F.,  {Rusli} S.~P.,  2007, \mnras, 374, 931

\bibitem[\protect\citeauthoryear{{Rieke} \& {Lebofsky}}{{Rieke} \&
  {Lebofsky}}{1985}]{1985ApJ...288..618R}
{Rieke} G.~H.,  {Lebofsky} M.~J.,  1985, \apj, 288, 618

\bibitem[\protect\citeauthoryear{{Ritchie}, {Clark}, {Negueruela} \&
  {Langer}}{{Ritchie} et~al.}{2010}]{Ritchie:2010lr}
{Ritchie} B.~W.,  {Clark} J.~S.,  {Negueruela} I.,    {Langer} N.,  2010, \aap,
  520, A48

\bibitem[\protect\citeauthoryear{{Ritchie}, {Clark}, {Negueruela} \&
  {Najarro}}{{Ritchie} et~al.}{2009}]{Ritchie:2009lr}
{Ritchie} B.~W.,  {Clark} J.~S.,  {Negueruela} I.,    {Najarro} F.,  2009,
  \aap, 507, 1597

\bibitem[\protect\citeauthoryear{{Salpeter}}{{Salpeter}}{1955}]{Salpeter:1955l%
r}
{Salpeter} E.~E.,  1955, \apj, 121, 161

\bibitem[\protect\citeauthoryear{{Skrutskie}, {Cutri}, {Stiening} \& {et
  al.}}{{Skrutskie} et~al.}{2006}]{Skrutskie:2006uq}
{Skrutskie} M.~F.,  {Cutri} R.~M.,  {Stiening} R.,    {et al.} 2006, \aj, 131,
  1163

\bibitem[\protect\citeauthoryear{{Spitzer}}{{Spitzer}}{1987}]{Spitzer:1987lr}
{Spitzer} L.,  1987, {Dynamical evolution of globular clusters, Princeton
  University Press, Princeton}

\bibitem[\protect\citeauthoryear{{Stetson}}{{Stetson}}{1987}]{Stetson:1987qy}
{Stetson} P.~B.,  1987, \pasp, 99, 191

\bibitem[\protect\citeauthoryear{{Stolte}, {Brandner}, {Brandl} \&
  {Zinnecker}}{{Stolte} et~al.}{2006}]{Stolte:2006kx}
{Stolte} A.,  {Brandner} W.,  {Brandl} B.,    {Zinnecker} H.,  2006, \aj, 132,
  253

\bibitem[\protect\citeauthoryear{{Stolte}, {Brandner}, {Grebel}, {Lenzen} \&
  {Lagrange}}{{Stolte} et~al.}{2005}]{Stolte:2005vn}
{Stolte} A.,  {Brandner} W.,  {Grebel} E.~K.,  {Lenzen} R.,    {Lagrange} A.,
  2005, \apjl, 628, L113

\bibitem[\protect\citeauthoryear{{Testi}, {Sargent}, {Olmi} \&
  {Onello}}{{Testi} et~al.}{2000}]{Testi:2000uq}
{Testi} L.,  {Sargent} A.~I.,  {Olmi} L.,    {Onello} J.~S.,  2000, \apjl, 540,
  L53

\bibitem[\protect\citeauthoryear{{Westerlund}}{{Westerlund}}{1961}]{1961PASP..%
.73...51W}
{Westerlund} B.,  1961, \pasp, 73, 51

\bibitem[\protect\citeauthoryear{{Whitmore}, {Chandar}, {Schweizer},
  {Rothberg}, {Leitherer}, {Rieke}, {Rieke}, {Blair}, {Mengel} \&
  {Alonso-Herrero}}{{Whitmore} et~al.}{2010}]{2010AJ....140...75W}
{Whitmore} B.~C.,  {Chandar} R.,  {Schweizer} F.,  {Rothberg} B.,  {Leitherer}
  C.,  {Rieke} M.,  {Rieke} G.,  {Blair} W.~P.,  {Mengel} S.,
  {Alonso-Herrero} A.,  2010, \aj, 140, 75

\bibitem[\protect\citeauthoryear{{Williams}}{{Williams}}{1999}]{Williams:1999l%
r}
{Williams} J.,  1999, in {J.~Franco \& A.~Carraminana} ed., Interstellar
  Turbulence, {The Structure of Molecular Clouds: are they Fractal?}.
p.~190

\bibitem[\protect\citeauthoryear{{Williams}, {Blitz} \& {McKee}}{{Williams}
  et~al.}{2000}]{Williams:2000fk}
{Williams} J.~P.,  {Blitz} L.,    {McKee} C.~F.,  2000, V. Mannings, A. P.
  Boss, S. S. Russell, eds., Protostars and Planets IV, University of Arizona
  Press, Tucson, p.~97

\end{thebibliography}
%\bibliographystyle{mn2e}

\appendix

\section{Completeness maps}
\label{app:compmaps}

In this appendix we will illustrate, step-by-step, how the 2D completeness maps for Wd~1 were obtained\footnote{The corresponding IDL procedures will be made available by the authors upon request.}.

\subsection{Adding and detecting artificial stars}
\label{subsec:artstars}
The basic idea is to use the same PSF that was obtained by PSF-fitting with DAOPHOT to add stars (using the DAOPHOT \texttt{addstar} task) in the reduced images and then run the same PSF fitting photometry scheme to see whether artificial stars can be recovered or not.
50 stars per run were added, in order not to change the crowding characteristic of the frame. Stars are positioned randomly on the frame and have an uniform distribution in magnitude. To achieve sufficient spatial resolution we iterate the procedure until we have added 4500 stars per unit magnitude. 
The actual size of the magnitude bin used is 0.5 mag, and we consider that artificial stars are recovered if the detection yields a difference between the input and output magnitude of less than 0.5 mag.
The ranges of magnitude spanned by the simulations are those typical of the Wd~1 member stars detected in the NTT/SofI observations, above the linear-regime limit and below the detection threshold i.e. $J \in [11.35, 19.35] \, \rmn{mag}$ and $K_\rmn{S} \in [9.8, 17.8] \, \rmn{mag}$; these magnitude ranges correspond to masses between 0.3 and 30 $M_{\sun}$, the exact values depending on the age, distance and extinction values (see Sect. \ref{sec:fundpar}).  The full magnitude ranges are  divided in 16 bins, 0.5 mag wide, per each of the two bands. 

With an effective detector area of $A_{\rmn{eff}} = L_x \times L_y = 876 \times 920 \, \rmn{pixel}^2 $, resulting from the area in common between the observations in the different filters, the numbers we just quoted correspond to a typical separation between simulated stars (within the 0.5 mag bin) of:
\begin{equation}
 \label{eq:sampdist}
 < d > = \sqrt{\frac{A_{\rmn{eff}}}{\upi \, N_{bin}}} \approx 10.8 \, \rmn{pixel}
\end{equation}
where $N_{bin} = 2225 $ is the number of simulated stars in the 0.5 mag-wide bin.
Given the SofI plate scale of $0\farcs29 / \rmn{pixel} $ and given the typical seeing of $\approx 0\farcs8$ it means that we sample the whole frame on a scale which is about 3.5 times the PSF Full-Width at Half Maximum (FWHM). The effective sampling scale is a bit larger, due to the use of a certain number of nearest neighbours to calculate the local value of completeness at the position of  each simulated star.
The natural limit, i.e. the minimum length-scale at which completeness can be sampled by our method, is the FWHM of the PSF itself, which characterizes the ability to distinguish two different point sources. The resolution of the incompleteness  map could not be improved further, even if the number of simulated stars would be increased in order to achieve a spatial sampling smaller than the PSF-FWHM. 
Our choice of the total number of stars is a compromise between a short sampling scale and a reasonable number of simulations.% To decrease the scale by the residual 3.5 factor, according to equation (\ref{eq:sampdist}), we should have simulated $(3.5)^2 \times 4500 = 55125$ stars per unit magnitude per each photometric band.

\subsection{Building the 2D maps}
\label{subsec:building}

Each simulated star can be either recovered or not by DAOPHOT PSF fitting, meaning that for that specific star completeness is either 0 or 1. Starting from this series of sparsely sampled 0-s and 1-s, several steps are necessary to obtain a smooth function, which is determined at each point on the frame. In the following we will indicate the position of simulated stars with a hat symbol, $(\hat{x},\hat{y})$, while the coordinate grid on which we actually calculated the function will be simply $(x,y)$, which corresponds to the pixels grid of the chip.
We will refer in this section only to a single magnitude bin and to a single photometric band; interpolation in the magnitude dimension will be treated in Sect. \ref{subsec:maginterp}.

The first step is to create average completeness values at each $(\hat{x}_i,\hat{y}_i)$ for $i=1, N_{bin}$. This is accomplished by considering a certain number $\nu$ of nearest neighbours to the i-th simulated star and defining the completeness fraction as:
\begin{equation}
\label{eq:Czero}
\mathcal{C}_0 (\hat{x}_i,\hat{y}_i) = \frac{\rmn{recovered \, stars}}{\nu + 1} \quad ;
\end{equation}
where the recovered stars are counted among the $\nu$ neighbours of the i-th one, which is also included, hence the +1 in the denominator.
The actual value of $\nu$ is somewhat arbitrary and has to satisfy two opposite requirements; the higher it is, the less the completeness values will be affected by statistical noise. On the other hand, a too large value would imply a loss in spatial resolution for our completeness maps. As mentioned in Sect. \ref{subsec:artstars} the effective sampling scale is not the $<d>$ of equation (\ref{eq:sampdist})
but more precisely:
\[
d_{\rmn{eff}} = <d> \times \sqrt{\nu} \quad .
\]

After several experiments we decided to use  $\nu = 16$, which degrades our completeness sampling scale by a factor 4, giving $d_{\rmn{eff}} \approx 43.2$ pixels, corresponding to about 14 times the FWHM of the image PSF.

At this stage the $\mathcal{C}_0$ is known only point-wise in the set of $(\hat{x},\hat{y})$ positions occupied by the simulated stars.
The next step is to interpolate this function into a regular grid of points. This is accomplished via the IDL procedure \texttt{GRIDDATA}, using the  Kriging  method of interpolation with an exponentially decreasing model for the variogram.
Kriging allows to interpolate a random field known in a set of positions into another set, under some assumptions about its covariance. In our case the random field is the completeness itself, with its Poissonian error due to the finite number of simulated stars considered in equation (\ref{eq:Czero}). An exponential model for the covariance is appropriate here, because the estimated values of $\mathcal{C}_0$ at some location $(\hat{x}_i,\hat{y}_i)$ are correlated with those for other stars and the correlation is stronger for closer simulated stars than for those separated by a large distance. We have chosen an $e$-folding scale equal to $d_{\rmn{eff}}$.

After the interpolation we performed a smoothing of the completeness. The grid used for the interpolation is indeed finer than $d_{\rmn{eff}}$, meaning that the interpolated function may show artificial variations on a scale smaller than our minimum size, which would be unrealistic. That is why we additionally smoothed the maps with a boxcar kernel with a size of $ d_{\rmn{eff}}$. The boxcar model is appropriate given the uniform spatial distribution of simulated stars.

\subsection{Interpolation in magnitude}
\label{subsec:maginterp}

Once the maps are available in magnitude layers, we enforced that at each location $(x,y)$, completeness is a decreasing function of magnitude. We fitted pixel-by-pixel a monotonically decreasing function of Fermi-like type:
\begin{equation}
 \label{eq:fermifun}
	\mathcal{C}_j (x, y, \umu) = \frac{\alpha (x,y)}{e^{\frac{\umu - \beta(x,y)}{\gamma(x,y)}}+1}
%	\mathcal{C}_j (x_l, y_l, \umu) = \frac{\alpha (x_l,y_l)}{e^{\frac{\umu - \beta(x_l,y_l)}{\gamma(x_l,y_l)}}+1}
\end{equation}
The meaning of the three coefficients is the following:
\begin{description}
 \item $\alpha$ is the normalization and is always $\leq 1$;
 \item $\beta$ is the magnitude for which completeness is $\alpha /2$;
 \item $\gamma$ represents the rapidity with which $\mathcal{C}_j$ drops down.
\end{description}
%From equation (\ref{eq:fermifun}) we have that 50\% completeness is reached for:
%\[
%\mu_{\rmn{half}} (x,y) = \beta + \gamma \times \log(2 \alpha - 1) 
%\]
Once the $(\alpha,\beta,\gamma)$ coefficients triplets are calculated, it is straightforward to assign to each real star its completeness value using the coefficients evaluated at the star's position.

\subsection{Completeness for the control field}
\label{subsec:compfield}

The offset field that we used for field decontamination of the colour-magnitude diagram (CMD) is also affected by incompleteness.
In this case, however, it is not necessary to investigate the 2D structure of the completeness pattern; under the assumption that the spatial distribution of the stars in the control field is uniform, we only consider spatially uniform incompleteness correction.

When using a control field for decontamination of a star cluster's CMD, one implicitly assumes that the stars are, on average, representative of the foreground/background population in the cluster's field. This assumption has a series of shortcomings. For example,  the copious cluster population itself may partially ''shield'' background stars. In addition to that, in the Galactic Plane, variable extinction may cause differences in the observed population of stars. Furthermore, the population along different lines of sight could be \emph{intrinsically} different, due to the different galactocentric distances sampled at the same heliocentric distance or changes within the spiral arms. This two latter problems are reduced by choosing nearby fields, so that the fore/background populations show  similar distributions in age and extinction - hence in magnitude and colour - as along the cluster's line of sight.
Hence the choice of the control field is done in order to have a population that \emph{on average} looks like the contaminating population in the cluster frame. 

For these reasons it is not necessary to try the same 2D approach to assign completeness values to the off-cluster frame stars. We only populated the whole frame in a uniform way with 250 stars per each 0.5 magnitude-wide bin; only 50 stars were added in each run, not to alter the crowding characteristics of the field. Then we computed the fraction of recovered stars over the total number of simulated ones and  fit a function like that of equation (\ref{eq:fermifun}), this time without any spatial dependence.
Last step was to assign also to the single stars in the control field their corresponding completeness value in each photometric band.

\section{Evaluation of the photometric errors and their correlations}
\label{app:everr}

In Paper I we used simulated stars to estimate the photometric errors. We showed that for stars with known input magnitudes the output magnitudes were often in disagreement at a level of more than $1\sigma$, where the  the DAOPHOT fitting errors were taken as  $\sigma$ values.
Hence, the difference between input and output magnitude seemed to be a more conservative and robust estimate of the real photometric error.
Simulated stars are used here also to estimate the correlation between magnitude errors.
The photon counts associated to an isolated star in two different bands are uncorrelated.
In reality, even though the photon counts are independent, the inferred magnitude values may not be.
The reason why $J \, \rmn{and} \, K_{\rmn{S}}$ magnitude errors are correlated is the presence of bright stars or, more generally, crowding.
When a faint star is located close to a bright star, the residuals of PSF fitting procedure of the bright star (which is usually bright in both bands) may lead to magnitude errors in both bands. If the bright star's wings are not properly subtracted, then there will be an excess in the flux that is assigned to the nearby faint star. The bright star's wings may also be over-subtracted (e.g. because the core is not well fitted), leading to too small flux estimates. This can lead to a correlation of the photometric errors. Crowding from stars of comparable magnitudes will lead to a similar behaviour.

\begin{figure}
 \centering
\resizebox{\hsize}{!}{\includegraphics{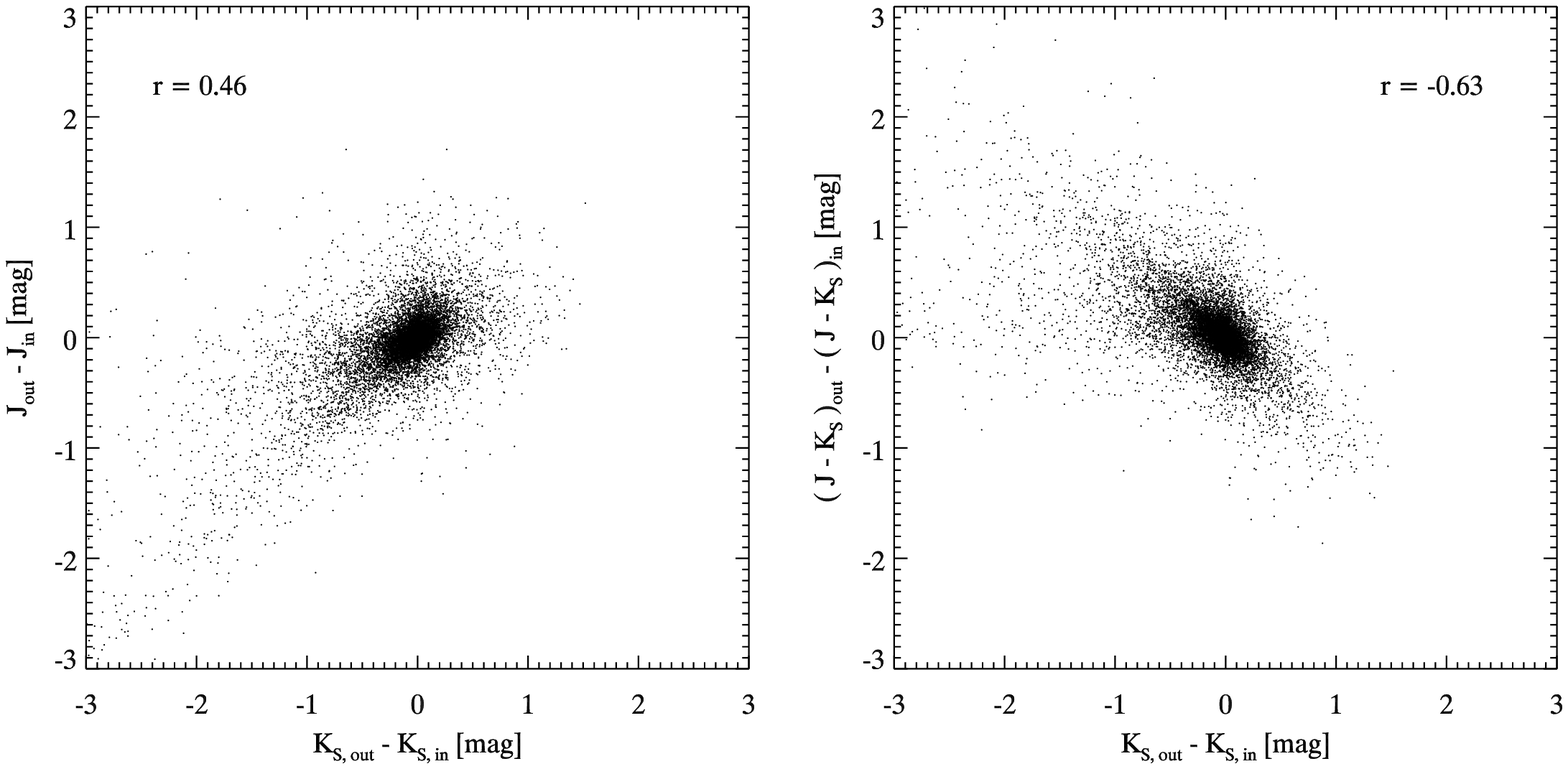}}
 \caption{Left: correlation between $J$ and $K_{\rmn{S}}$ magnitude errors for the simulated stars. Right: same, but using $(J - K_{\rmn{S}})$ colour instead of $J$ magnitude. The $r$ quantities are the values of Pearson's correlation coefficients for the two samples.}
 \label{fig:corr}
\end{figure}

That $J$ and $K_{\rmn{S}}$ magnitude errors  are correlated is obvious from Fig. \ref{fig:corr}. In the left panel we show $J^{out} - J^{in} \,\rmn{vs.} \,K^{out} - K^{in}$ for the simulated stars. In the right panel we show $(J - K_{\rmn{S}})^{out} - (J - K_{\rmn{S}})^{in} \,\rmn{vs.}\, K^{out} - K^{in}$. Since the two magnitude estimates are correlated, the composed quantity $J - K_{\rmn{S}}$ is also correlated to the single magnitude values. The coefficient $r$ in the figures is the Pearson's correlation coefficient for the whole sample of simulated stars, i.e.
\begin{equation}
\label{eq:Pearson}
   r_{X,Y} = \frac{ \sum_i {\left(X_i - \overline{X}\right)\left(Y_i - \overline{Y}\right)} }{ \sqrt{ \sum_i{\left(X_i - \overline{X}\right)^2}} \, \sqrt{ \sum_i{ \left(Y_i - \overline{Y}\right)^2}} } \quad ; 
\end{equation}
where X and Y are the respective abscissa and ordinate in the two plots.
From its definition it is clear that Pearson's $r$ is equivalent to the X and Y covariance divided by the product of X and Y standard deviations.
A value of $r$ very close to +1 (-1) indicates a very tight correlation (anticorrelation) between the two random variables, while two totally uncorrelated random variables would show a value of $r=0$.
The values of the Pearson's coefficients in Fig. \ref{fig:corr} indicate a quite significative correlation of the magnitude errors as well as an even tighter anticorrelation between magnitude and colour errors.
Given that $r$ is not a robust, outlier-resistant quantity, the actual values were calculated removing the outliers, i.e. stars for which input and output magnitudes differ more than 1 mag in at least one band . The number of stars inside this limits is 97 \% of the total number of simulated stars. Therefore, the exclusion of the outliers does not represent a shortcoming in the evaluation of a robust estimate for the overall correlation coefficient of the sample.

\subsection{Assigning proper photometric errors and their correlation to each detected star}
\label{subsec:errorassign}
\begin{figure}
 \centering
\resizebox{\hsize}{!}{\includegraphics{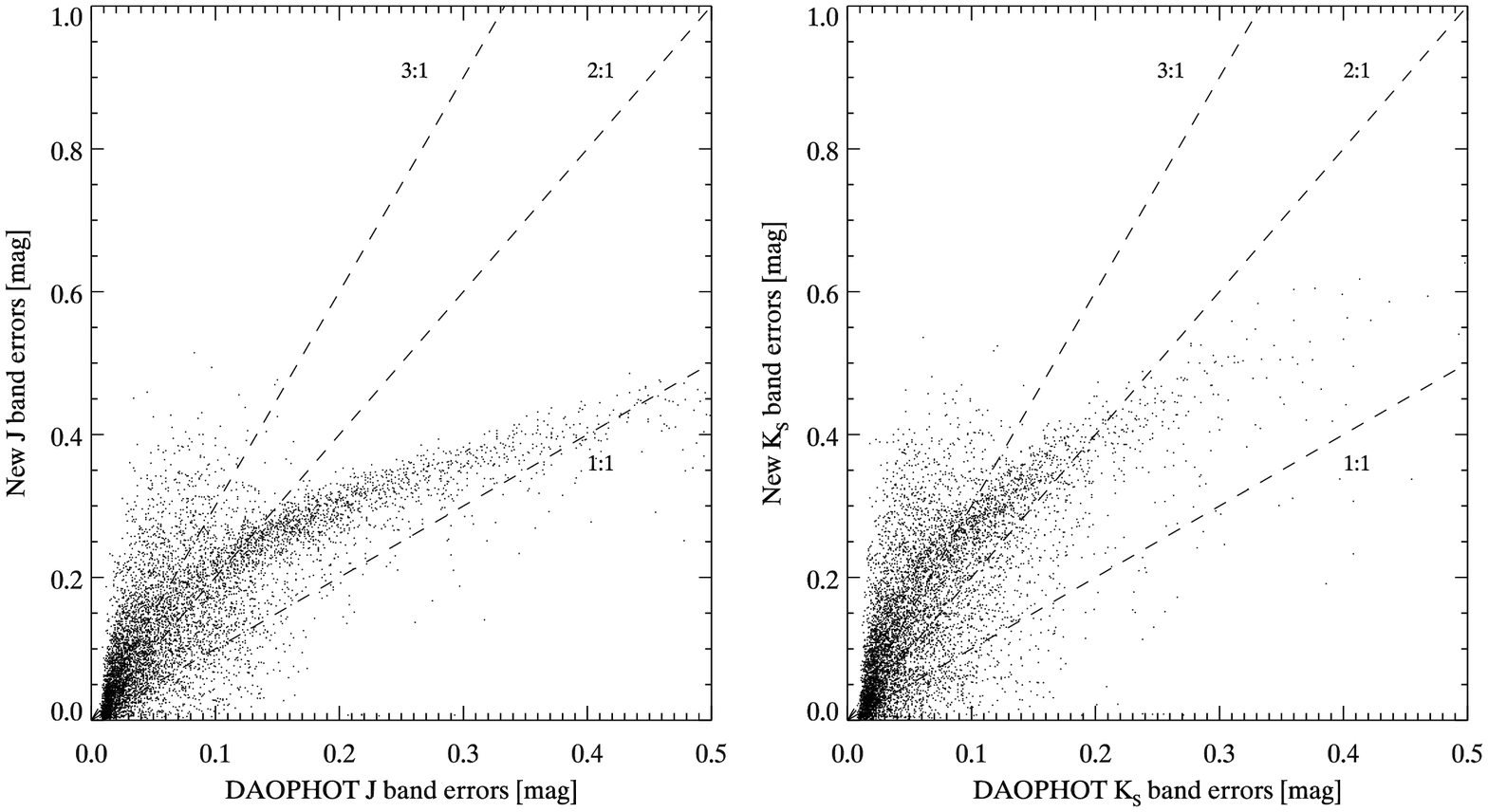}}
 \caption{Comparison between DAOPHOT photometry errors and our newly derived errors for $J$ band (left) and $K_{\rmn{S}}$ band in the case of Wd~1 field. The dashed lines are 1:1, 2:1 and 3:1 relations between new and DAOPHOT errors.}
 \label{fig:newolderr}
\end{figure}

For each detected star we selected a number of at least 7 simulated stars (using the same stars of Sect. \ref{subsec:artstars}) that were positioned in its \emph{neighbourhood}. By neighbours we mean simulated stars whose distance from the position of the real star was not larger than 50 pixels and whose magnitude differs no more than 1 magnitude - in each band - from the real star. The distance of  50 pixels represents the average \emph{radius-of-influence} of the bright stars, i.e. the typical extent of their halos, as derived by the analysis of Wd~1 NTT/SofI images.
For each of the neighbours in the subsample we calculated $J^{out} - J^{in} \,\rmn{and} \,K^{out} - K^{in}$.
The standard deviations of the two quantities, within the subsample, have been used as estimates for the photometric errors of the real stars.
We also calculated the Pearson coefficient between the two quantities in the neighbours subsample and assigned it to the detected star.
The comparison between DAOPHOT errors and our newly estimated errors for the Wd~1 field are shown in Fig. \ref{fig:newolderr}. The new error estimates are, on average, larger than what predicted by DAOPHOT, especially in the $K_{\rmn{S}}$ band.

Some of the real stars do not have a sufficient number of neighbours to perform this kind of estimate. This is true especially for faint stars, since the majority of the simulated stars at the faintest magnitudes cannot be recovered. Hence these simulated stars cannot be used for the error estimate, because they don't have a $M^{out}$ value. For stars without enough useful simulated neighbours, we used a different error estimate. We first divided the real stars for which the error determination worked fine in several magnitude bins. Then we calculated the mean error per each magnitude bin and fitted an exponential relation to the \emph{mean error} vs. \emph{bin-magnitude} points. This relation has been used to assign their errors to the stars that lack a sufficient number of neighbours. 
The new errors, as a function of the stars' magnitudes are shown in Fig. \ref{fig:newerrvsmag}. The figure shows the exponential extrapolation used to determine the errors of the faintest stars.
Together with the new errors for the cluster's field stars, we show, in red, the new errors for the control field stars, whose derivation is illustrated in Sect. \ref{subsec:photerrref}.

\subsection{Photometric errors for the control field}
\label{subsec:photerrref}
A similar method was used to derive new photometric errors for  the control field stars. Since we assume that these stars are uniformly distributed, there is no need to treat the spatial variation of the errors. Using the same simulated stars as in Sect. \ref{subsec:compfield}, we computed the $\Delta M_j (i) = M_j^{out} (i) - M_j^{in} (i)$ for  $M_j =J, K_{\rmn{S}}$ and $i$ running over the simulated stars; then we binned the stars in 0.5 mag wide bins (in the input magnitudes) and for each bin we computed the standard deviation of the $\Delta M_j$ over the bin.
The last step was to fit the $\left( \sigma [\Delta M_j] , M_j [\rmn{bin}] \right)$ points with an exponential relation; here $M_j [\rmn{bin}]$ is the central magnitude of the bin. This relation was used to assign an error to the real stars  as a function of magnitude.
The average value of the correlation between $J$ and $K_{\rmn{S}}$ was calculated for the whole sample and is $ r = 0.25$. This value was assigned to each real star in the control field.

Figure \ref{fig:newerrvsmag} shows that, on average, the photometric errors in the control field are smaller than those in the cluster's field. This behaviour is expected, and can be explained by the higher degree of crowding for Wd~1's field. For the same reason, the detection limit for the control field is $\sim 0.5 \,\rmn{mag}$ fainter than the Wd~1 field in both photometric bands.

\section{$\sigma$ - clipping}
\label{subsec:sigmaclip}

Because of some dissimilarities between the on- and off-field fore/background populations, the CMD of Wd~1, even after subtraction, does not look perfectly clean. For this reason, after having chosen the best fitting isochrone, i.e. the 4 Myr one, before any further analysis, we additionally subtracted those stars that lie more than $3 \sigma$ away from the reference isochrone in the magnitude-magnitude space (see the end of Sect. \ref{sec:subfield}).
After clipping, essentially all stars with colours and magnitudes consistent with the 4 Myr cluster population are included in the  final source selection.
Our clipping criteria may retain some arbitrariness; nevertheless they do not affect our further analysis. The main reason is the cut only affects the faint stars, with large photometric errors. Some of them could be excluded or included in the catalogues by slightly changing the $\sigma$ threshold. Anyway, in the computation of the IMF (see Sect. \ref{sec:IMF}) and of the stellar density (see Sect. \ref{sec:morph}), we only consider stars above a given \emph{completeness} or \emph{mass} threshold. Stars with uncertain membership are mostly excluded by these two additional cuts hence they do not affect the final results.

One realization of the clean cluster's CMD is shown in Fig. \ref{fig:CMDclean}, together with the best fitting isochrone. The error bars shown in the diagram are the average $J-K_{\rmn{S}}$ and $K_{\rmn{S}}$ errors per magnitude bin.
The colour errors are calculated for each star as:
\begin{equation}
\label{eq:colerr}
 \delta(J-K_{\rmn{S}}) = \sqrt{  \sigma^2(J) + \sigma^2(K_{\rmn{S}}) + 2 r_{JK_{\rmn{S}}} \sigma(J) \sigma(K_{\rmn{S}}) }
\end{equation}
Pearson's $r$ is equal to the covariance divided by the product of the two standard deviations (see equation \ref{eq:Pearson}); hence the third addend on the r.h.s. of equation (\ref{eq:colerr}) is equal two twice the covariance of $J$ and $K_{\rmn{S}}$.

In Table \ref{tab:starnumbers} we summarize the number of stars left  after field subtraction and additional clipping.
The mean values and their uncertainties are derived by iterating the probabilistic subtraction technique. We repeated the extraction of $\zeta$ for each star, to generate 100 different catalogues (see Sect. \ref{sec:subfield}). We then calculated mean and standard deviations of the number of members over the 100 samples.

\setcounter{table}{0}
\begin{table}
 %\centering
\begin{center}
% use packages: array
 \caption{Detections in the on and off frames.}
 \label{tab:starnumbers}
\begin{tabular}{ll}
\hline
Field & Number of stars \\
\hline
 on         &          7036           \\
 off         &          5381           \\
 on (after subtraction)         &  $5810 \pm 25$                   \\ 
on  (after subtraction and $\sigma$-clipping)       &          $4300 \pm 23 $           \\
\end{tabular}
\end{center}
\end{table}

\section{Bootstrap estimate of the IMF parameters and their errors}
\label{subsec:IMFpar}

Bootstrapping is a resampling technique for error estimation \citep[see e.g.][]{Efron:1979lk,Hastie2009ji,Andrae:2010lr}. Given a data set from which some parameters are estimated, bootstrapping consists in resampling the data  to create \emph{alternative} data sets. From these, it is possible to repeatedly estimate the parameters of interest, monitoring their distribution.
We generated $10^5$ bootstrap samples to probe the parameter space of $(\gamma,A)$, assuming for the IMF the functional form $\frac{\rmn{d}N(m)}{\rmn{d}m} = \mathcal{A} \times m^{ - \gamma}$.
From our data set, we created 100 different realizations of the members catalogue. Each catalogue has a slightly different number of members $N_{c,j} \,\rmn{with}\, j=1,100$, after statistical field subtraction and $\sigma$-clipping (see Appendix \ref{subsec:sigmaclip}).
From the members of each $j$-th catalogue 1000 bootstrap samples were created. The new samples consist of the same number of stars as in the member catalog, $N_{c,j}$, but the drawing is made with replacement, i.e., the same star can occur multiple times in a bootstrap sample.
This sample of stars is then used to build the IMF as in equation (\ref{eq:IMFdef}) where now $i$ runs on the stars of the specific bootstrap sample.
At each iteration a power-law fit is performed to obtain a couple $(\gamma_{j,k},A_{j,k}) \,\rmn{with}\,j=1,100 \,\rmn{and}\, k=1,1000$.
As already detailed in Sect.\ref{subsec:IMFdet}, the fitting interval is restricted to $m\in[3.5,27] \,M_{\sun}$.
Given the $(\gamma_{j,k},A_{j,k})$ values, we obtained the corresponding total mass, $M_{j,k}$, and total number of stars, $N_{j,k}$, by integrating the power law in the interval $m\in[0.08,120 M_{\sun}]$.

\begin{figure}
 \centering
\resizebox{\hsize}{!}{\includegraphics{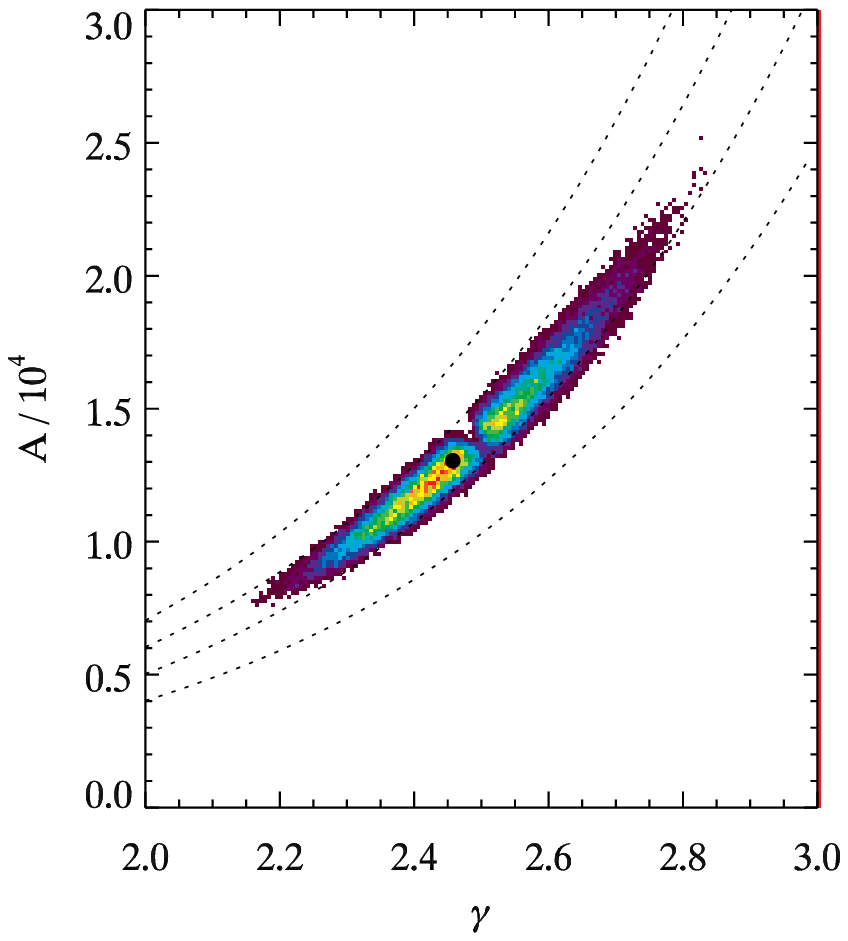}}
 \caption{Density of the occurrencies of $(\gamma, A)$ from our bootsrap procedure. The black dot indicates the location of the maximum of the 2D density: $(\gamma_{\rmn{M}}, A_{\rmn{M}}) = (2.46, 1.31\times10^4)$. Dotted lines are lines of constant number of stars in the fitting interval; from bottom to top $N_{\rmn{fit}} =$ 1000, 1250, 1500, 1750.}
 \label{fig:densboot}
\end{figure}

A 2D density plot of the output values $(\gamma,A)$ is shown in Fig. \ref{fig:densboot}.
It is clear that the $\gamma$ and $A$ parameter are tightly correlated. This is easy to understand. For each bootstrap sample we have a number $N_{\rmn{fit}}$ of stars that are actually inside the fitting interval. Given the different catalogue realizations, this number can be slightly different, but is mostly in the interval [1250, 1500]. The IMF fit has to satisfy the condition:
$$
N_{\rmn{fit}} = A \times \int_{3.5}^{27} m^{-\gamma} \rmn{d}m 
$$ 
From which we get:
$$
A = \frac{N_{\rmn{fit}}(1-\gamma)}{27^{1-\gamma}-3.5^{1-\gamma}}
$$
This relation between $A$ and $\gamma$ is overplotted  in Fig. \ref{fig:densboot} for $N_{\rmn{fit}} =$ 1000, 1250, 1500, 1750 (dotted lines).
Given that the 2D distribution of $(\gamma,A)$ pairs is clearly non Gaussian, the definition of the best values and the confidence intervals for the two parameters is not straightforward.
The maximum of the 2D distribution is located at $(\gamma_{\rmn{max}}, A_{\rmn{max}}) = (2.46, 1.31\times10^4)$. With this pair of values we obtain a total mass, $M_{max} = 5.13 \times 10^4 \, M_{\sun}$ and a total number $N_{max}= 1.10 \times 10^5$ stars.

\begin{figure*}
 \centering
\resizebox{\hsize}{!}{\includegraphics{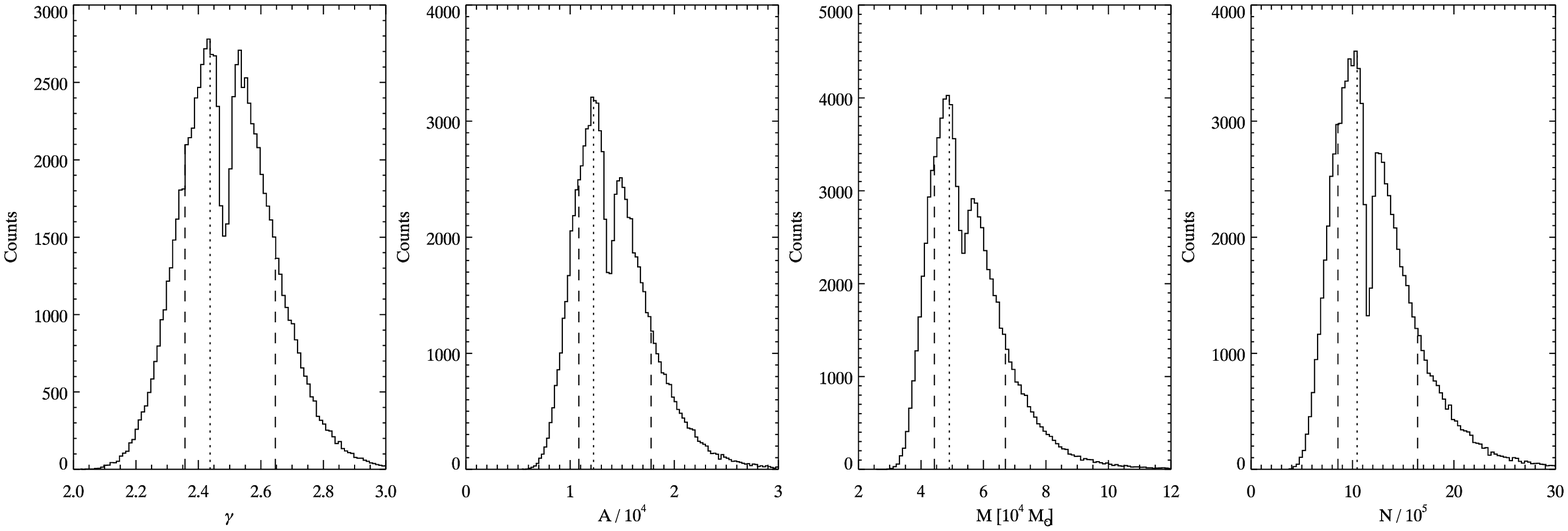}}
 \caption{From left to right: Marginal distributions for $\gamma$ and $A$ and distributions of $M_{j,k}$, and $N_{j,k}$. The best values are marked by dotted lines; the confidence intervals are marked by dashed lines.}
 \label{fig:margdistr}
\end{figure*}

On the other hand, using the 2D joint distribution is not the most suitable choice for defining the best values and confidence interval for the parameters $(\gamma, A)$ and for $M_{\rmn{tot}}$ and $N_{\rmn{tot}}$.
For this purpose, in the case of $\gamma$ and $A$, we used the marginal distributions. These are obtained by integration of  the joint distribution with respect to the other variable. For $M_{\rmn{tot}}$ and $N_{\rmn{tot}}$ we similarly used the distributions of $M_{j,k}$, and $N_{j,k}$ obtained after each bootstrap iteration.
The best values are obtained by maximizing the distributions. The confidence intervals are obtained by integrating the distributions from left and right until 16\% of the total area under the distribution is reached on each side. This means that the limits of the asymmetric confidence interval comprise 68\% of the total area under the distribution function. The marginal distributions for $\gamma$ and $A$, as well as the distributions of $M_{j,k}$, and $N_{j,k}$ are shown in Fig. \ref{fig:margdistr}. The best values and the confidence intervals are given in Table \ref{tab:bootout}.
% use packages: array

\setcounter{table}{0}
\begin{table}
 %\centering
\begin{center}
% use packages: array
 \caption{Best values and their confidence intervals for the IMF parameters, the total mass and the total number of stars of Wd~1..}
 \label{tab:bootout}
\begin{tabular}{lccc}
\hline
Quantity & Best value & Lower limit & Upper limit \\
\hline
 $\gamma$         &          2.44        &    2.36             &  2.64  \\
 $A / 10^4$       	&        1.22          &   1.08            & 1.78 \\
 $M_{\rmn{tot}} [10^4 M_{\sun}]$         &          4.91        & 4.42                &  6.70  \\
 $N_{\rmn{tot}} / 10^4$         &       10.4       &   8.6              &  16.4  \\

\end{tabular}
\end{center}
\end{table}

\label{lastpage}

\end{document}